\documentclass[a4paper,11pt]{article}
\usepackage{jcappub}
\usepackage{amsmath,empheq}
\usepackage[x11names]{xcolor}
\usepackage{hyperref}
\gdef\@fpheader{ }
\makeatother
\DeclareMathOperator\arctanh{arctanh}
\DeclareMathOperator\Chie{Chi}

\DeclareMathOperator\Ci{Ci}

\title{On the constant roll complex scalar field inflationary models}

\author[a]{Ali Mohammadi,}
\author[a,1]{Nahid Ahmadi,\note{Corresponding author.}}
\author[a,b]{Mehdi Shokri}

\affiliation[a]{Department of Physics, University of Tehran, Kargar Ave. North, Tehran 14395-547, Iran}
\affiliation[b]{School of Physics, Damghan University, P. O. Box 3671641167, Damghan, Iran}

\emailAdd{alimohammadi95@ut.ac.ir}
\emailAdd{nahmadi@ut.ac.ir}
\emailAdd{mehdishokriphysics@gmail.com}

\begin{document}

\abstract{
In this paper we wish to point out the possibility of using a complex scalar field in a constant roll inflationary model, as needed for observational viability. We extend the idea of real field inflaton with constant rate of roll to a complex field, showing the feasibility of solving Einstein Klein-Gordon equations constrained  by an \emph{appropriate} form of constant roll definition. As compared to the well known (two-parametric class of) real field models, there is one more degree of flexibility in constant roll inflationary solutions which is represented by an arbitrary function of time, $\gamma(t)$. We work with an arbitrary but constant function $\gamma$ (where $\gamma=0$ refers to the corresponding real field model) and find new inflationary class of potentials. In this class of models, the behavior of real and complex field models are similar in some aspects, for example the solutions with large constant roll parameter are not stable and should be considered as early time transients. These field solutions relax at late time on a dual attractor trajectory. However, complex fields phase space trajectories reach this stable regime after real fields. We performed the stability analysis on $\gamma$ function space solutions and found that dynamically stable trajectories in phase space are stable under $\gamma$ variations. We extended this study by considering multifield models of constant roll inflation with non-canonical kinetic terms. By enlarging the size of field space, we showed that a multifield constant roll model is dynamically a single field effective theory. If field space is parametrized by $N$ non-canonical fields, there will be $N$ free parameters in the potential that can be attributed to the interaction between the fields. 
 }
\maketitle

\section{Introduction}\label{sec:introduction}

There is no doubt that inflation is an important part of modern cosmology \cite{STAROBINSKY198099, Sato:1980yn, Guth:1980zm, PhysRevLett.48.1437, Linde:1981mu, Lyth:1998xn, Martin:2016ckm}. In its simplest realizations, an accelerated expansion is driven by a single real scalar field $\phi$, called an inflaton. The dynamical system of equations that describe this period of time includes the Klein-Gordon equation and Friedmann equation. It is generally not possible to solve this system; so standard assumptions like slow roll, in which $\ddot{\phi}\ll H\dot{\phi}$, are made to simplify the dynamics. This assumption yields a simple realization with viable observational predictions; nevertheless, the increase in accuracy of cosmological data over the recent years has been a motivator to go beyond this assumption and generalize the inflation driven by slow rolling field. The cosmological generalizations have been studied either from GR or particle physics point of view \cite{Hazra:2014jka,Kinney:2005vj,Nojiri:2017qvx,Motohashi:2017vdc,Oliveros:2019zkl,delCampo:2012qb,Kehagias:2018uem,Das:2020xmh,Liu:2021diz,Yuennan:2022zml,Palti:2019pca, Andriot:2018mav, Talebian-Ashkezari:2016llx, Firouzjahi:2020jrj, Firouzjahi:2018vet}. An inventive generalization from the latter viewpoint, was developed by Motohashi et. al. \cite{Motohashi:2014ppa}. This is a two-parametric class of models in which the inflaton is assumed to roll with constant rate. The novelty is that by introducing this assumption to the dynamical system, it is $\it{possible}$ to solve the dynamical system exactly and construct the new type of inflation dubbed constant roll (CR) inflation. A considerable number of variants for this model can be found in the literature, some of them are \cite{Awad:2017ign, Motohashi:2019tyj, Odintsov:2019ahz,Micu:2019fju,Gao:2019sbz,Guerrero:2020lng,Shokri:2021rhy,MohseniSadjadi:2019vvs, Oikonomou:2021yks,Shokri:2021jxh,Shokri:2021aum,Shokri:2021iqp,Shokri:2021zqw,Stein:2022cpk, Anari:2022tsl, Herrera:2023ywx}.

In constant roll construction, inflaton was assumed to be a real scalar field. A complex scalar field has higher degree of freedom that makes its predictions more flexible to match with data. Such fields have been invoked in many different areas of physics. A minimally coupled complex field yields the same field equations as those obtained by two scalar fields. Cosmological scenarios with such fields are extensively studied~\cite{Yurov:2001ud,Yurov:2002nu,Buchmuller:2014epa,Carrion:2021yeh, Scialom:1996yd,Gu:2001tr,Liu:2020bmp,Amendola:1994xf,Scialom:1994uq,Khalatnikov:1992sj,Kamenshchik:1997dmk,Kamenshchik:1995ib} and in the present work we revisit the idea focusing on the constant roll assumption. 
The inflaton potential in real field models does not require the parameter defining the constant roll be small. It is important to see the degree of generality of this possibility.

As an assumption, not every interpretation of constant-roll assumptions would be consistent with the dynamical system.
 As a dynamical system a relevant question is: Under what constraints (imposed on the value or velocity of fields) a complex field constant roll inflation can be realized? For a real field the constant roll potential is completely fixed; so another relevant question is whether the complex field potentials are also fixed. What about the a field space with larger dimension? 

The paper is arranged as follows. In sec.~\ref{sec:Background Dynamics} we begin by a short review of a complex field dynamics and then examine couple of traditionally used CR constraints and discuss the possibility of finding inflationary potential. We also derive a sufficient condition on the potential to ensure the compatibility in the constrained dynamical system. In Sec.~\ref{CR Potential}, we find explicit solutions for the field, Hubble parameter and the field potential. This introduces a three-parametric class of solutions that satisfies the constrained system exactly. The subject of (non-)uniqueness of the solutions is also discussed. The dynamical behavior of the fields' solution is the subject of Sec.~\ref{Dynamical Analysis}, where the stability of the parameter space solution are also analyzed. Sec.~\ref{multifield} is devoted to the analysis of CR multifield models with non-canonical kinetic terms i.e with a curved field space. The approach in earlier sections, will be generalized to show its robustness in a field model with arbitrary dimension of fields. We summarize our results and conclude in Sec.~\ref{Conclusion}. The paper ends with one appendix where the constrained system is formulated in a field space with canonical coordinate and exact potential solution is derived.
\section{Constant Roll Dynamical System}\label{sec:Background Dynamics}

In this section, we present a detailed analysis of an expanding Universe with a complex field as the field responsible for the accelerating expansion. We review the Einstein Klein-Gordon equations as a dynamical system and then discuss about the compatibility of a constraint which may be imposed on this system.

\subsection{Einstein Klein-Gordon equations}\label{Einstein Klein-Gordon equations}
We consider a spatially flat FLRW background described by
\begin{equation}
ds^2=-dt^2+a^2(t)d\vec{x}^2.
\end{equation}
The action for the Universe with a minimally coupled complex field is
\begin{equation}
S=\int{\sqrt{-g}d^4 x\left(\frac{R}{2}+\frac{1}{2}g^{\mu\nu}\left(\partial_\mu \Phi^\ast\right)(\partial_\nu \Phi)-V\left(\Phi\right)\right)}.
\end{equation}
One can represent the field degrees of freedom (DoFs) by the amplitude $X\left(t\right)$ and the phase $\theta \left(t\right)$ or $\Phi=\varphi_1(t)+i \varphi_2(t)$. The Lagrangian density can be written in canonical or non-canonical form
\begin{eqnarray}
\mathcal{L}_\Phi&=&-\frac{1}{2}a^3 \left(\dot{\varphi_1}^2+\dot{\varphi_2}^2\right)+a^3 V\left(\varphi_1,\varphi_2\right)\label{Lagrangian phi1-phi2}\\
&=&-\frac{1}{2}a^3 X^2\left[\left(\frac{\dot{X}}{X}\right)^2+\dot{\theta}^2\right]+a^3 V\left(X,\theta\right).
\end{eqnarray}
The variation of the action in the non-canonical form yields the Einstein and the complex field equations of motion as
\begin{gather}
3H^2=\frac{1}{2}\left(\dot{X}^2+X^2\dot{\theta}^2\right)+V(X,\theta),\label{Friedmann1}\\
-2\dot{H}=\dot{X}^2+X^2\dot{\theta}^2,\label{Friedmann2}\\
{\ddot{X}}-\dot{\theta}^2 X+3H \dot{X}+\frac{\partial V}{\partial X}=0,\label{KG-X}\\
X^2\ddot{\theta}+3HX^2\dot{\theta}+2X\dot{X}\dot{\theta}+\frac{\partial V}{\partial \theta}=0.\label{KG-theta}
\end{gather}
For $\dot{\theta}=0$ and $V=V(X)$, the standard equations for a real scalar field is recovered. In real field case, the only field equation is the differential consequence of Friedmann equations. The above set of equations for complex field are not independent as well. In this case, The Bianchi identity guarantees that equations ~\eqref{Friedmann1},~\eqref{Friedmann2} and~\eqref{KG-X} automatically imply ~\eqref{KG-theta} and the above system of equations is also fully determined by the independent equations ~\eqref{Friedmann1} -~\eqref{KG-X}. By taking the time derivative of ~\eqref{Friedmann1} and using ~\eqref{Friedmann2}, we find
\begin{equation}\label{Friedmann time derivative}
\dot{X}\left[{\ddot{X}}-\dot{\theta}^2 X+3H \dot{X}+\frac{\partial V}{\partial X}\right]+\dot{\theta}\left[X^2\ddot{\theta}+3HX^2\dot{\theta}+2X\dot{X}\dot{\theta}+\frac{\partial V}{\partial \theta}\right]=0.
\end{equation}
Therefore any solution of Friedmann equation will imply one of fields equations (~\eqref{KG-X} or~\eqref{KG-theta}), if the other one is already satisfied. In other words, If in the set of comprising Einstein plus two Klein-Gordon equations, one of Klein-Gordon equations be a linear combinations of the other, the system will be under-determined. In such case, both Klein-Gordon equations can be neglected and the dynamical system shall be fully described by Einstein equations. 

If $\partial V/\partial\theta=0$, the phase variable $\theta$ would be cyclic and there would be one independent DoF; so~\eqref{KG-theta} can be solved to give $\dot{\theta}={M}/({a^3 X^2})$, where $M$ is a constant. Clearly, this system would be different from a real field, because if the centrifugal terms in kinetic energy and the field equation is expressed in terms of the conserved quantity $M$, and scale factor $a$, one gets the following set of equations
\begin{equation}
3H^2=\frac{1}{2}\left(\dot{X}^2+\frac{M^2}{a^6 X^2}\right)+V(X),  \qquad {\ddot{X}}-\frac{M^2}{a^6 X^3} +3H\dot{X}+\frac{d V}{d X}=0.
\end{equation}
 Note, however, that in real field case, there is no explicit dependency on the scale factor in the Klein-Gordon equation.

 In the rest of this section, we will study the dynamical system described in this section by imposing the constant roll constraint. Before embarking on that study, we review the constant roll definitions for multi-field inflationary models.
\subsection{Constant Roll Definitions}\label{sec:CR definitions}
The inflationary models with an approximately flat potential yield a sufficiently long period of quasi-de Sitter expansion and a nearly scale invariant spectrum of density perturbations. However, there has been considerable interest to find non-standard inflationary exact solutions to the equations of motion. We now come to the constraints which may be imposed on the dynamical equations (\ref{Friedmann1}-\ref{KG-theta}) to $\it{define}$ a class of non-slow roll inflationary scenarios with constant rates of roll. Starting with a single real field $\phi$, the assumption of a constant rate of roll is formulated by
\begin{equation}\label{CR constraint1}
\ddot{\phi}=-\eta H\dot{\phi}.
\end{equation}
Here $\eta$ is a constant parameter that is equivalent to the second slow-roll parameter for values that describe a dynamically stable system. The standard slow-roll regime occurs at $\eta\simeq0$, while the ultra-slow-roll case corresponds to $\eta=-3$. Although data seem to favor the small values of $\eta$ \cite{Motohashi:2014ppa,Motohashi:2017aob}, but the novelty of CR class of models is that the exact inflationary potential can be found without need to consider $\eta$ as a small parameter.\footnote{In Sec.~\ref{Dynamical Analysis} we will talk about a duality relation which suggests that the ultra slow-roll inflationary models may also be consistent with the observations.} One may think of generalizations of this constraint when more degrees of freedom play role in an inflationary scenario. When a complex field is responsible for the CR inflation, three trivial generalizations may be used for
\begin{itemize}
	\item Both DoFs constant roll independently, i.e,
	\begin{equation}\label{CR constraint 2}
  \ddot{X}=-\eta H\dot{X}, \qquad \ddot{\theta}=-\eta H\dot{\theta}.
\end{equation}
\item The constant rate of roll is conjectured only for absolute value of the field velocity,
\begin{equation}\label{CR constraint 21}
\frac{d|\dot{\Phi}|}{dt}=-\eta H|\dot{\Phi}|.
\end{equation}
\item The definition ~\eqref{CR constraint1} has an equivalent form, written in terms of the Hubble parameter,
\begin{equation}\label{CR constraint3}
\ddot{ H}=-2\eta H\dot{H}.
\end{equation}
\end{itemize}

Equation~\eqref{CR constraint3} \emph{seems} to be independent of the matter dynamics and therefore believed to be more fundamental than~\eqref{CR constraint1}. We shall see in next section that the second definition~\eqref{CR constraint 21} can be derived from this form. So, the latter form practically affects on the matter dynamics. It has been used for realizing a constant roll inflation in string theory \cite{Micu:2019fju}. We will see that the definition~\eqref{CR constraint3} fixes the time profile of potential, but there remains some flexibility when expressed in terms of DoFs. In the next subsection we elaborate on these definitions and show that it is actually impossible to construct an exact solution, if the definition~\eqref{CR constraint 2} is added to the system of equations. We also discuss about the consistency condition between the CR definition~\eqref{CR constraint3} and other equations in dynamical system discussed in subsection~\ref{Einstein Klein-Gordon equations}.
\subsection{Consistency Condition}\label{sec: consistency condition}
As we discussed in Subsection \ref{Einstein Klein-Gordon equations}, unlike a real field model, the inflationary potential cannot be determined from Einstein equations plus CR constraint. For a complex field, we show that such a set of equations is considered overdetermined and almost always inconsistent.

Starting with first CR definition, it can be written in the form of $\ddot{X}/\dot{X}=\ddot{\theta}/\dot{\theta}=-\eta H$. This can be integrated to give
\begin{equation}\label{X and theta relation}
X=M\theta+M_0.
\end{equation}
Here $M$ and $M_0$ are integration constants with mass dimension. An expression for the potential is found by applying~\eqref{X and theta relation} to the first Friedmann equation~\eqref{Friedmann1},
\begin{equation}\label{V1}
V\left(X,\theta(X)\right)=3H^2-\frac{1}{2}(1+\frac{X^2}{M^2})\dot{X}^2.
\end{equation}
Now let us apply~\eqref{X and theta relation} to dynamical equations in~\eqref{Friedmann time derivative}. We will have
\begin{equation}\label{Friedmann time derivative2}
\dot{X}\left[{\ddot{X}}+3H \dot{X}-\frac{1}{M^2}\dot{X}^2 X+\frac{\partial V}{\partial X}\right]+\frac{X^2\dot{X}}{M^2}\left[\ddot{X}+3H\dot{X}+2\frac{\dot{X}^2}{X}+\frac{M^2}{X^2}\frac{\partial V}{\partial X}\right]=0.
\end{equation}
It is easy to see that the two squared brackets do not vanish simultaneously. In other words, the potential given in~\eqref{V1} does not necessarily satisfy both field equations. One therefore concludes that the above definition of CR complex field model is not mathematically consistent.

A system like above, in which the equations outnumber the unknowns, is overdetermined and almost always inconsistent. It will, however, have solutions in some cases. For imposing CR constraint to the dynamical system one can consider the conditions which render the system under-determined, i.e, some equations are linear combinations of the others. We carry on this section to find a condition to render the field equations dependent. Let us concentrate on kinetic energy and 
the relative contribution of different DoFs at each time. We define a smooth function related to the kinetic energy function,
\begin{equation}\label{Z definition}
Z\equiv\dot{X}^2+X^2\dot{\theta}^2,
\end{equation}
in terms of which dynamical equations (\ref{Friedmann1}-\ref{KG-theta}) are given by
\begin{equation}\label{Friedmann3}
3H^2=\frac{1}{2}Z+V, \qquad -2\dot{H}=Z.
\end{equation}
 If the contribution of different DoFs in $Z$ be proportional to each other at all times, i.e, $X\dot{\theta}=\lambda\dot{X}$, with $\lambda$=constant, we will get $Z=(1+\lambda^2)\dot{X}^2$ and different DoFs are related by
 \begin{equation}\label{theta(X)}
\theta-\theta_0=\lambda \ln(X).
\end{equation}
The special case of real field is recovered by $\lambda=0$ and $\theta(\lambda=0)=\theta_0$, which is a constant. Now let us apply the relation between DoFs~\eqref{theta(X)}, to~\eqref{KG-theta}, we obtain
\begin{equation}\label{KG-theta2}
\lambda X\left(\ddot{X}+3H\dot{X}\right)+\lambda \dot{X}^2+\frac{\partial V}{\partial \theta}=0.
\end{equation}
One strategy is to find a relation between partial derivatives of potential to guarantee that both equations~\eqref{KG-X} and~\eqref{KG-theta} be satisfied simultaneously. We, therefore, replace the expression $\ddot{X}+3H\dot{X}$ in~\eqref{KG-theta2} by its value obtained from~\eqref{KG-X} and get
 \begin{equation}\label{consistency condition}
\lambda X\frac{\partial V}{\partial X}-\frac{\partial V}{\partial \theta}=\lambda (1+\lambda^2)\dot{X}^2.
\end{equation}
The $\lambda=0$ case trivially holds and the extension of above discussion to $\lambda=\lambda(t)$, adds an extra term $X\dot{X}\dot{\lambda}$ to the right hand side of~\eqref{consistency condition}. In the following, we consider~\eqref{consistency condition} as the consistency condition that $V(X,\theta)$ must satisfy.
\section{A Consistent CR Potential}\label{CR Potential}
Following the discussion in subsection \ref{sec: consistency condition}, a universal definition for CR multi-field models that can be applied to complex fields is~\eqref{CR constraint3}. Based on this definition, the time profile of Hubble parameter and scale factor are given by
\begin{equation}\label{H(t)}
H(t)=C_1\frac{C_2 e^{C_1\eta t}+e^{-C_1\eta t}}{C_2 e^{C_1\eta t}-e^{-C_1\eta t}}, \qquad  a(t)=C_3\left(C_2 e^{C_1\eta t}-e^{-C_1\eta t}\right)^{1/\eta}.
\end{equation}
 Here $C_i$s, $(i=1,2,3)$ are integration constants and $C_1$ represents a mass scale. Mathematically,~\eqref{H(t)} gives a solution of~\eqref{CR constraint3} for any complex values of $C_i$s. We, however, focus on the cases with real values. For potential $V(t)$, we get
\begin{equation}\label{V(t)}
V(t)=3 H^2+\dot{H}= C_1^2 \left(3+\frac{4 (3-\eta) C_2}{\left(C_2 e^{C_1 \eta t}-e^{- C_1 \eta t} \right)^2} \right) .
\end{equation}
Substituting the time derivative of the first relation in~\eqref{H(t)} into~\eqref{CR constraint3}, the CR definition can be written in this new equivalent form
\begin{equation}\label{CR constraint4}
\dot{Z}=-2\eta HZ.
\end{equation}
Recalling that $Z=|\dot{\Phi}|^2$, the equivalence of CR definitions~\eqref{CR constraint 21},~\eqref{CR constraint3} and~\eqref{CR constraint4} is evident. Furthermore, from $Z=-2\dot{H}$ and the Hubble parameter expression in~\eqref{H(t)}, it is easy to see that
\begin{equation}\label{Z(t)}
Z(t)=\frac{8\eta C_1^2 C_2}{\left(C_2 e^{C_1\eta t} - e^{-C_1 \eta t}\right)^2}.
\end{equation}
Obviously, to have a positive kinetic energy, we need $Z>0$. This immediately implies that the parameter $\eta$ and the integration constant $C_2$ have the same sign.

By $Z(t)$ in hand, we use the following ansatz $\dot{X}=-\sqrt{Z}\cos(\gamma)$ and $X\dot{\theta}=\sqrt{Z}\sin(\gamma)$ to determine the contribution of different DoFs in kinetic energy, at time $t$.\footnote{Note that for $\cos(\gamma)>0$, the minus sign for $\dot{X}$ is consistent with the real field inflation model.}
Here $\gamma\in\left(-\frac{\pi}{2},\frac{\pi}{2}\right)$ is an arbitrary function of time and by choosing any specific function for $\gamma(t)$, a relation between DoFs shall be established. The main source of this arbitrariness is the fact that in a system with different DoFs, the map between $t$ and DoFs is not generally one-to-one; so the procedure of replacing the time parameter in~\eqref{V(t)} by a function of $X$ and $\theta$ is quite arbitrary. This arbitrariness can be seen in another way in an equivalent form of equation~\eqref{CR constraint4} given by
\begin{equation}\label{CR constraint5}
X^2\dot{\theta}\ddot{\theta}+\dot{X}\ddot{X}+X\dot{X}\dot{\theta}^2=-\eta H(\dot{X}^2+X^2\dot{\theta}^2).
\end{equation}
To pinpoint the CR evolution of DoFs, the above form can be decomposed in different ways. Some simple decompositions are
\begin{subequations}\begin{empheq}[]{align}
&\ddot{X}=-\eta H\dot{X} \qquad X\ddot{\theta}=-\eta H X\dot{\theta}-\dot{X}\dot{\theta}, \label{decomposition 1}\\
&\dot{X}\ddot{X}+X\dot{X}\dot{\theta}^2=-\eta H\dot{X}^2 \qquad \ddot{\theta}=-\eta H\dot{\theta} \label{decomposition 2}.
\end{empheq}\end{subequations}
Either decomposition corresponds to a particular relation between DoFs, for example $X=X(\theta(t),t)$. One can talk about a single-valued map $X(t)$ or $\theta(t)$ iff the decomposition gives $X(t)=X(\theta(t))$. These expressions may be inverted to give $t(X)$ and $t(\theta)$ as well as other dynamical quantities like Hubble parameter, $\dot{\theta}$ or $\dot{X}$ in terms of $X$ or $\theta$.

In the following, we work with~\eqref{decomposition 1} that constraints the absolute value of the complex field by $\ddot{X}=-\eta H\dot{X}$. By imposing the above ansatz into~\eqref{Friedmann3}, we find that $\dot{X}\propto X\dot{\theta}\propto\exp\left(-\eta\int{H dt}\right)$. This yields a relation between DoFs similar to~\eqref{theta(X)}, with $\lambda=\tan(\gamma)=$constant. We recall that by choosing $\gamma=$constant the contribution of each DoF in kinetic energy will be fixed, for all times. In this simple case, the exact expression for $X(t)$ and $\theta(t)$ are given by
 \begin{subequations}\begin{empheq}[]{align}
&X(t)=-\cos(\gamma)\int{\sqrt{Z(t)}dt}\propto
\cos(\gamma)\arctanh\left(\sqrt{C_2} e^{C_1 \eta t}\right),\label{X(t)}\\
&\theta(t)=\sin(\gamma)\int{\frac{\sqrt{Z(t)}}{X(t)}dt}=\tan(\gamma)\int{\frac{\sqrt{Z(t)}}{\int^t{\sqrt{Z(\acute{t})}d\acute{t}}}dt}=\tan(\gamma)\ln\left(\frac{X(t)}{\cos(\gamma)}\right).\label{theta(t)}
  \end{empheq}\end{subequations}
 Although the potential time profile~\eqref{V(t)} is fixed by this CR definition, we discuss how $V(X,\theta)$ can be found, if we work with~\eqref{decomposition 1}. Let us consider the potential in a sum separable form $V=V_1\left(X\right)+V_2(\theta)$, and then find $dV_1/dX$ and $dV_2/d\theta$ from the field equations~\eqref{KG-X} and~\eqref{KG-theta}. After employing the decomposition given in~\eqref{decomposition 1}, we will have
 \begin{subequations}\begin{empheq}[]{align}
&V_1=\int{\frac{dV_1}{dX}}dX=\int{(\eta-3)}H\dot{X}dX+\int{X\dot{\theta}^2 dX },\label{V_1}   \\
&V_2=\int{\frac{dV_2}{d\theta}d\theta}= \int{X\dot{\theta}\left[(\eta-3)H X-\dot{X}\right] d\theta}.\label{V_2}
\end{empheq}\end{subequations}
The real field potential is obtained from the first integral in~\eqref{V_1}. Furthermore, it is easy to check that the sum $V_1+V_2$ obtained from equations in~\eqref{V_1} and~\eqref{V_2}, satisfies the consistency condition~\eqref{consistency condition}. Although a more general form of a sum separable potential, i.e, $V=a V_1\left(X\right)+ b V_2(\theta)$, with constant positive definite values for $a$ and $b$, could have been considered, we take the simpler form $a=b=1$, for the consistency condition to be satisfied. Now, to have the exact expression for potential, we apply this procedure to two interesting model groups: 1) $\eta > 0$ and $C_2>0$ and 2) $\eta < 0$ and $C_2<0$. Note that  the condition $\eta C_2>0$ is necessary when CR definition~\eqref{CR constraint3} is used.
\subsection{$\eta > 0$ and $C_2>0$}\label{eta positive}
For $C_2>0$ case, one can define $\alpha=\frac{1}{2}\ln{C_2}$ and rewrite solutions~\eqref{H(t)} as
\begin{equation}\label{H(t) eta positive}
H(t)=C_1 \coth\left(C_1 \eta t+\alpha\right), \qquad a(t)=\tilde{C}_3 \sinh^{\frac{1}{\eta}}\left(C_1 \eta t+\alpha\right).
\end{equation}
To avoid singular behavior when $t\rightarrow 0$, we keep $C_2\neq 0$. Moreover, to ensure the positivity of Hubble parameter, $H>0$, we have to take $\left(C_1 \eta t+\alpha\right)\geq 0$ and $C_2>1$. Using~\eqref{X(t)} and following the discussion given in \cite{Anguelova:2017djf} about $\arctanh$ with an argument larger than one, we find that
\begin{equation}\label{X(t), eta positive}
X=\frac{1}{\kappa}\ln\left(\coth{\frac{1}{2}(C_1\eta t+\alpha)}\right),
\end{equation}
where a dimensionless parameter $\kappa:=\frac{\sqrt{2\eta}}{2\cos(\gamma)}$ is defined. As far as $t$ as a function of DoFs is concerned, one can invert~\eqref{X(t), eta positive} and get
\begin{equation}
\coth(C_1\eta t+\alpha)=\cosh\left(\kappa X\right)=\cosh\left(\sqrt{\frac{\eta}{2}}e^{\frac{\theta}{\tan(\gamma)}}\right).
\end{equation}
 In the second equality, equation~\eqref{theta(t)} is used. The Hubble parameter, $H$, and the kinetic energy related function, $Z$, expressed in terms of DoFs are given by 
\begin{eqnarray}\label{}
H=C_1 \cosh\left(\kappa X\right)=C_1 \cosh\left(\sqrt{\frac{\eta}{2}} e^{\frac{\theta}{\tan(\gamma)}}\right),\\
 Z=2\eta C_1^2\sinh^2\left(\kappa X\right)=2\eta C_1^2\sinh^2\left(\sqrt{\frac{\eta}{2}} e^{\frac{\theta}{\tan(\gamma)}}\right).
\end{eqnarray}
This, together with the ansatz $\dot{X}=-\sqrt{Z}\cos(\gamma)$ and $X\dot{\theta}=\sqrt{Z}\sin(\gamma)$ immediately gives $\dot{X}$ and $\dot{\theta}$. Now, we have whatever needed to do the integrations in~\eqref{V_1} and~\eqref{V_2}. We will finally get
\begin{eqnarray}\label{V(X,theta) eta positive}
V(X,\theta)&=&V_1(X)+V_2(\theta)\nonumber\\
&=& \text{constant}+\frac{{C_1^2}}{2}\left\{(3-\eta)\cos^2(\gamma)\cosh\left(2\kappa X\right)-2\eta \sin^2\left(\gamma\right)\left[\Chie\left(2\kappa X\right)-\ln(\kappa X)\right]\right.\nonumber\\
&+ & \left.\sin^2(\gamma)\left[{(3-\eta)}\cosh\left(\sqrt{2\eta}e^{\frac{\theta}{\tan(\gamma)}}\right)+{2\eta} \Chie\left(\sqrt{2\eta}e^{\frac{\theta}{\tan(\gamma)}}\right)-{2\eta} \ln\left(\sqrt{\frac{\eta}{2}}e^{\frac{\theta}{\tan(\gamma)}}\right)\right]\right\}.\nonumber\\
\end{eqnarray}
Here $\Chie(x)$ is the hyperbolic cosine integral defined by
\begin{equation}\label{Chie}
\Chie(x)=\gamma+\ln(x)+\int^x_0{\frac{\cosh(t)-1}{t} dt}.
\end{equation}
A comparison between~\eqref{V(t)} and~\eqref{V(X,theta) eta positive} gives the constant term equal to $\frac{C_1^2}{2}\left(3+\eta\right)$. It can also be found from the $\gamma\rightarrow 0$ limit of~\eqref{V(X,theta) eta positive}.

\subsection{$\eta< 0$ and $C_2<0$}\label{eta negative}
Let us now consider negative values of constant roll parameter. For these models, we substitute $C_2=-|C_2|$ and $\eta=-|\eta|$ in~\eqref{H(t)} and find
\begin{equation}\label{H1(t)}
H(t)=-C_1 \tanh(C_1|\eta| t+\beta),       \qquad a(t)=\bar{C}_3\cosh^{\frac{1}{\eta}}(C_1|\eta| t+\beta).
\end{equation}
 Here $\beta=-\frac{1}{2}\ln|C_2|$. In these models $t$ is limited to the range $(-\infty,-\frac{\beta}{C_1 |\eta|}]$. Recalling that $\arctanh(i x) =i\arctan(x)$ $\forall x$, we get
\begin{equation}\label{X(t), eta negative}
X=\sqrt{\frac{8}{|\eta|}}\cos(\gamma)\arctan\left(e^{C_1 |\eta| t+\beta}\right)=\frac{2}{\mu}\arctan\left(e^{C_1 |\eta| t+\beta}\right).
\end{equation}
Here $\mu=\frac{\sqrt{2|\eta|}}{2\cos(\gamma)}$ and we assumed $X(-\infty)=0$. Besides, we have
\begin{equation}
\exp(C_1 |\eta| t+\beta)=\tan\left(\frac{\mu}{2}X\right)=\tan\left(\sqrt{\frac{|\eta|}{8}} e^{\frac{\theta}{\tan(\gamma)}}\right).
\end{equation}
Like the previous case, Hubble parameter $H$ and $Z$ are easily found
\begin{eqnarray}
H&=&C_1\cos\left(\mu X\right)=C_1\cos\left(\sqrt{\frac{|\eta|}{2}} e^{\frac{\theta}{\tan(\gamma)}}\right),\\
Z&=&2|\eta|C_1^2\sin^2\left(\mu X\right)=2|\eta|C_1^2\sin^2\left(\sqrt{\frac{|\eta|}{2}} e^{\frac{\theta}{\tan(\gamma)}}\right),
\end{eqnarray}
and potential is given by
\begin{eqnarray}\label{V(X,theta) negative eta}
V(X,\theta)&=&V_1(X)+V_2(\theta)\nonumber\\
&=& \text{constant}+\frac{{C_1^2}}{2}\left\{(|\eta|+3)\cos^2(\gamma)\cos\left(2\mu X\right)+2|\eta|\sin^2(\gamma)\left[\ln(\mu X)-\Ci(2\mu X)\right]\right.\nonumber\\
&+&\left.\sin^2(\gamma)\left[(|\eta|+3)\cos\left(\sqrt{2|\eta|}e^{\frac{\theta}{\tan(\gamma)}}\right)+2|\eta|\left(\Ci\left(\sqrt{2|\eta|}e^{\frac{\theta}{\tan(\gamma)}}\right)-\ln\left(\sqrt{\frac{|\eta|}{2}}e^{\frac{\theta}{\tan(\gamma)}}\right)\right)\right]\right\}.\nonumber\\
\end{eqnarray}
Here, $Ci(x)$ stands for the cosine integrals, defined by\footnote{In the definitions~\eqref{Chie} and~\eqref{Ci} $\gamma$ is Euler–Mascheroni constant and should not mixed up with $\gamma$ used throughout this paper to determine the contributions of DoFs in kinetic energy.}
\begin{equation}\label{Ci}
\Ci(x)=\gamma+\ln(x)+\int^x_0{\frac{\cos(t)-1}{t} dt}.
\end{equation}
In a similar way, the constant term can be found to be $\frac{C_1^2}{2}(3-|\eta|)$.

\section{Dynamical Analysis}\label{Dynamical Analysis}

\begin{figure}\begin{center}
\includegraphics[scale=.25]{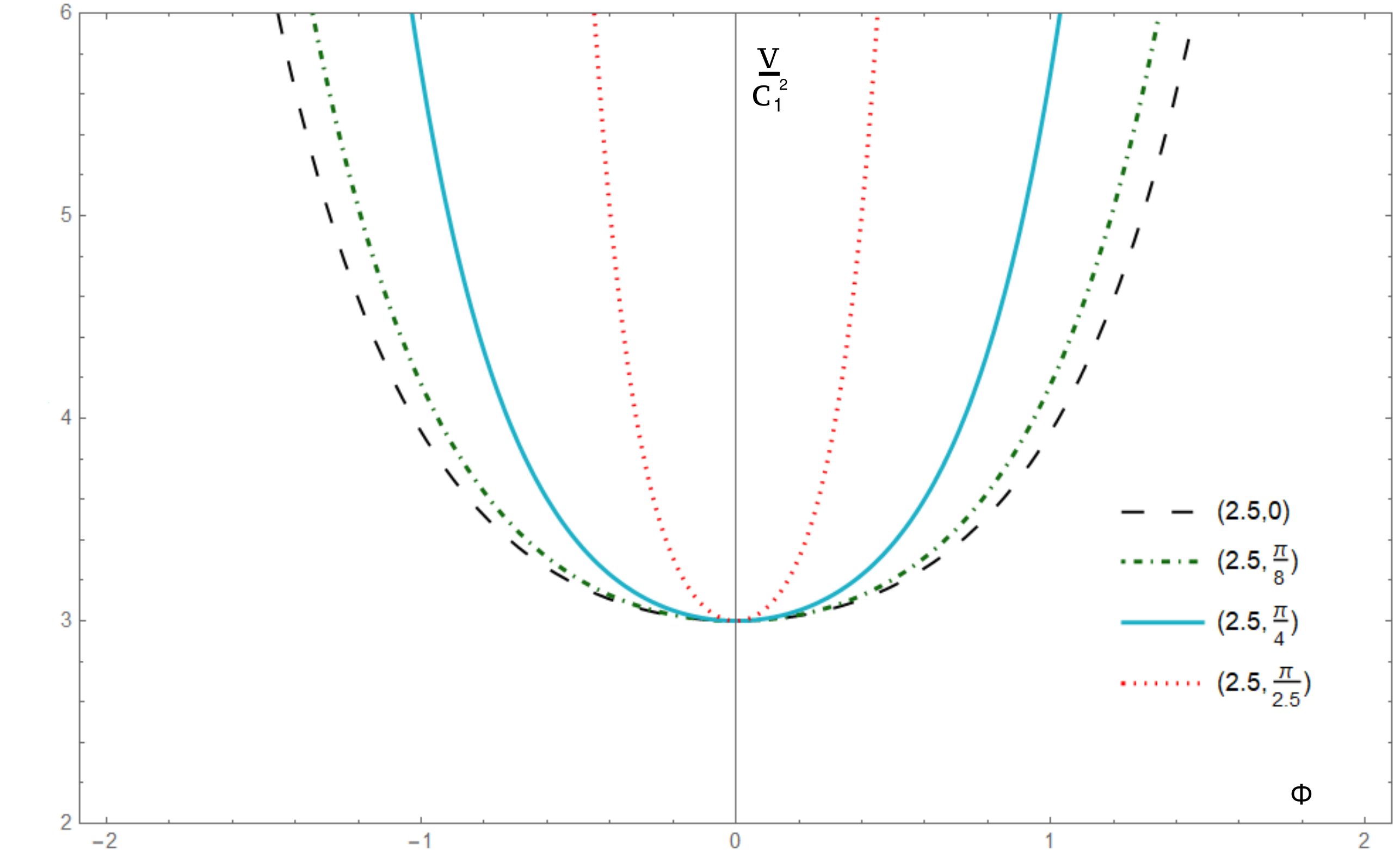}
\includegraphics[scale=.25]{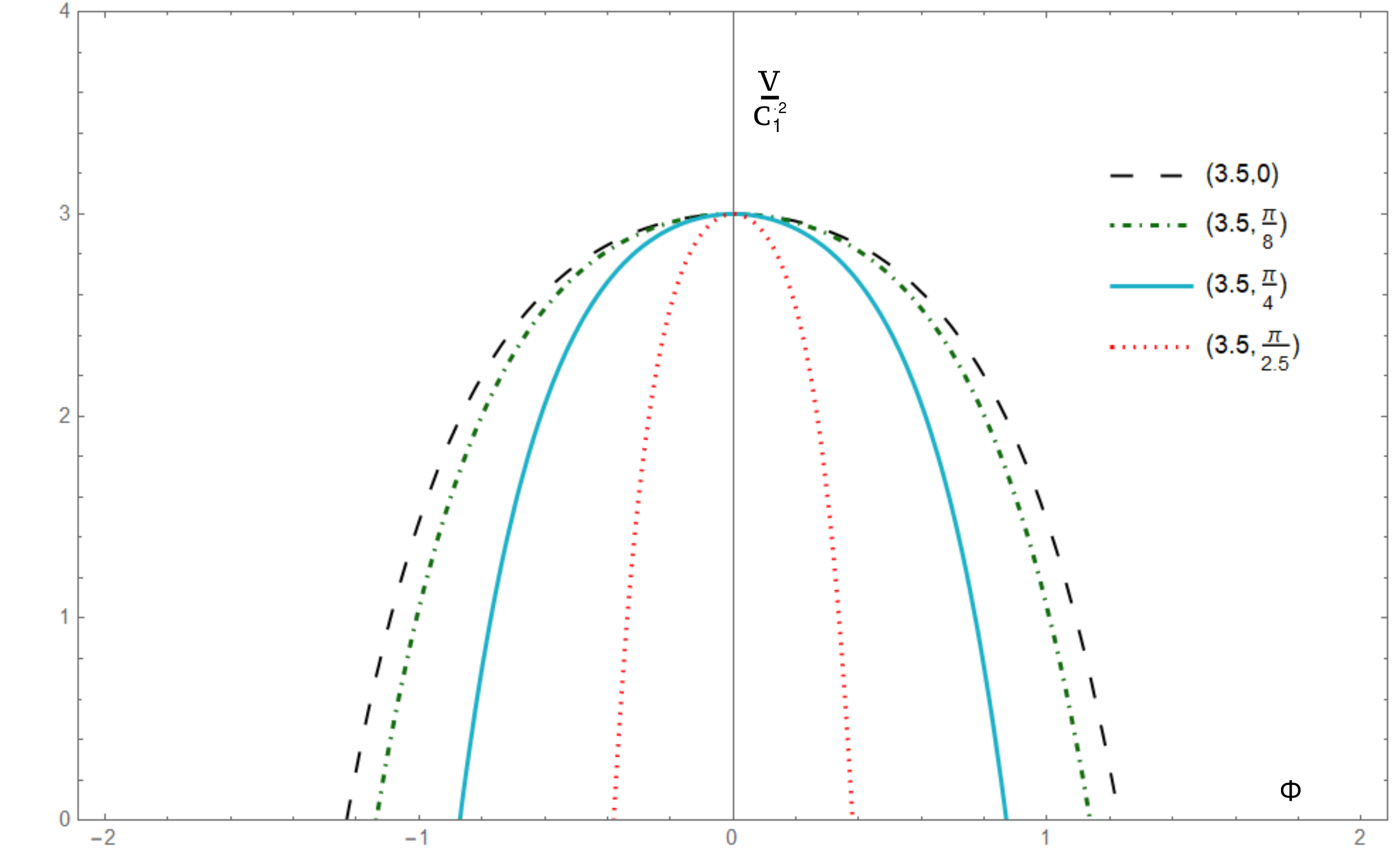} \end{center}
\caption{The form of potentials~\eqref{positive} for $\eta=2.5$ and $\eta=3.5$ and different values of $\gamma$ are compared with that of $\gamma=0$. The domain of validity of these potentials is limited to the right half of this plots and the other half is given for the comparison with $\gamma=0$.}%
\label{fig:potential plots}
\end{figure}

By combining equation~\eqref{KG-X} and~\eqref{KG-theta}, we get the dynamics of field in terms of the field absolute value\footnote{In the rest of paper we use different notations $X$, $|\Phi|$ or even $x$ equivalent to each other.} $|\Phi|$,
\begin{equation}\label{KG4}
|\dot{\Phi}|\frac{d|\dot{\Phi}|}{dt}+3H|\dot{\Phi}|^2+\frac{d V(|\Phi|)}{dt}=0.
\end{equation}
One can easily find more familiar form of equation~\eqref{KG4}, after applying the ansatz $\frac{d|\Phi|}{dt}=-\cos(\gamma)|\dot{\Phi}|$, as
\begin{equation}
\label{KG41}\frac{d^2|\Phi|}{dt^2}+(3H+\dot{\gamma} \tan(\gamma))\frac{d|\Phi|}{dt}+\cos^2(\gamma)\frac{dV(|\Phi|)}{d|\Phi|}=0.
\end{equation}
The form of potential solution found in previous subsection would be best described by $V(X,\theta)=V(X,\theta(X))$. 
  If the relation between DoFs is applied to either expression (~\eqref{V(X,theta) eta positive} and ~\eqref{V(X,theta) negative eta}), one can find $V(|\Phi|)$, as follows
\begin{subequations}\label{eqn:V(Phi)}\begin{empheq}[left={V(|\Phi|)= \empheqlbrace}]{align}
  &\frac{C_1^2}{2}\left[(3+\eta)+(3-\eta)\cosh\left(2\kappa |\Phi|\right)\right] \quad &\eta > 0, \label{positive} \\
  &\frac{C_1^2}{2}\left[(3-|\eta|)+(3+|\eta|)\cos\left(2\mu |\Phi|\right)\right] \quad &\eta < 0. \label{negative}
\end{empheq}  \end{subequations}
\label{eq:system}
These potentials have three fixed parameters $(C_1,\eta,\gamma)$. The first parameter determines the inflation scale. The qualitative form of potential depends on the values of $(\eta,\gamma)$ which is invariant under $\gamma\leftrightarrow -\gamma$. A plot of this can be seen in figure \ref{fig:potential plots}.
Generally, these forms are very much like the potentials found in the literature for real field CR models. For $0<\eta<3$, the potential is a convex function like a hybrid-type inflation, where a kind of transition is needed to end the inflation. For $\eta>3$, the potential is a concave function like a hilltop-type inflation. For $\eta<0$, the potential is again a concave function and is a particular type of hilltop inflation models. Here, the additional DoF has increased the slope of the potential and rescaled the field absolute value. The slope change, which is controlled by either $\kappa$ or $\mu$, is a natural consequence of partitioned kinetic energy. Figure \ref{fig:potential plots} shows that by increasing $\gamma$ the region around the extrema which can be approximated by a quadratic expression is reduced. The size of this region is important in the evolution of the system under the duality relation discussed in the literature \cite{Tzirakis:2007bf, Morse:2018kda,Gao:2019sbz}. $\eta=3$ is the critical value, in which the potential is constant and invariant under $|\Phi|\rightarrow |\Phi|+$constant.

In these particular solutions, in which $\eta$ is assumed to be constant, the inflation seems to take place near the maximum for concave potentials and ends as the field rolls down and in the case of convex potentials, the field  asymptote to rest at the bottom of the potential as $|\dot{\Phi}|\rightarrow 0$. To check whether this solution is an attractor or not we will concentrate on the phase space diagrams directed by these potentials, in the following subsection. On the other hand, the parameter $\gamma$ quantifies the contribution of additional DoF in kinetic energy and study of its evolution determines whether there are regimes that the phase of complex field affects the field dynamics. We therefore study the variations of $\gamma$, when the field has reached the attractor path in subsection~\ref{Stability Analysis}.
\subsection{Attractor Behavior-Phase Space Analysis}\label{Attractor Behavior}

Although the potentials given in~\eqref{V(X,theta) eta positive},~\eqref{V(X,theta) negative eta} and~\eqref{V(phi1,phi2)} look different, all three satisfy the following phase space equations 
\begin{equation}\label{non-autonomous equations}
\frac{d x}{dt}=-\cos(\gamma(t))y, \qquad  \frac{d y}{dt}=-3 H y-\frac{1}{y}\frac{d V}{d t}.
\end{equation}
Here we worked with dimensionless variables: $x\equiv|\Phi|$, $y\equiv|\dot{\Phi}|/C_1$, $t\equiv C_1 t$, $H\equiv H/C_1$, $V\equiv V/C_1^2$ and functions $H(t)$ and $V(t)$ are given by~\eqref{H(t)} and~\eqref{V(t)}, respectively. No matter what the arbitrary function $\gamma(t)$ is given, these non-autonomous nonlinear equations are common between all CR inflationary models which satisfy~\eqref{CR constraint4}. This universality is the result of choosing a CR definition which is independent of individual dynamics of DoFs.

The analysis of the inflationary dynamics for $\gamma=\gamma(t)$ is difficult, if not impossible to do. We, therefore, concentrate on the potentials given in~\eqref{eqn:V(Phi)}, in which $\gamma$ was assumed to be a constant and come back to $\gamma$ variations in next subsection. Using~\eqref{positive}, we will have a set of autonomous equations given by
\begin{equation}\label{autonomous equations}
\frac{d x}{dt}=-\cos(\gamma)y, \qquad \frac{d y}{dt}=-3 \cosh(\kappa x) y+(\eta-3)\sqrt{2\eta}\sinh(2\kappa x).
\end{equation}
We assume $C_2=1$, without any loss of generality. We have plotted the phase space trajectories for different values of $\gamma$ and $\eta=2.5,3.5$ in figure \ref{fig:(gamma,eta)}.  In the first row there is an attractor trajectory toward the minimum for the convex potential, whereas for the concave potential plots (second row) there is not. In second row plots we see trajectories that field velocity vanishes before passing the maximum of the potential where the field rolls back. For the trajectories that reach the maximum the field velocity $|\dot{\Phi}|$ may well be nonzero. This is the characteristic feature of a non-slow-roll system since in slow-roll limit $|\dot{\Phi}|\propto V'(|\Phi|)$ which vanishes at extrema. 

\begin{figure}
\includegraphics[scale=.1]{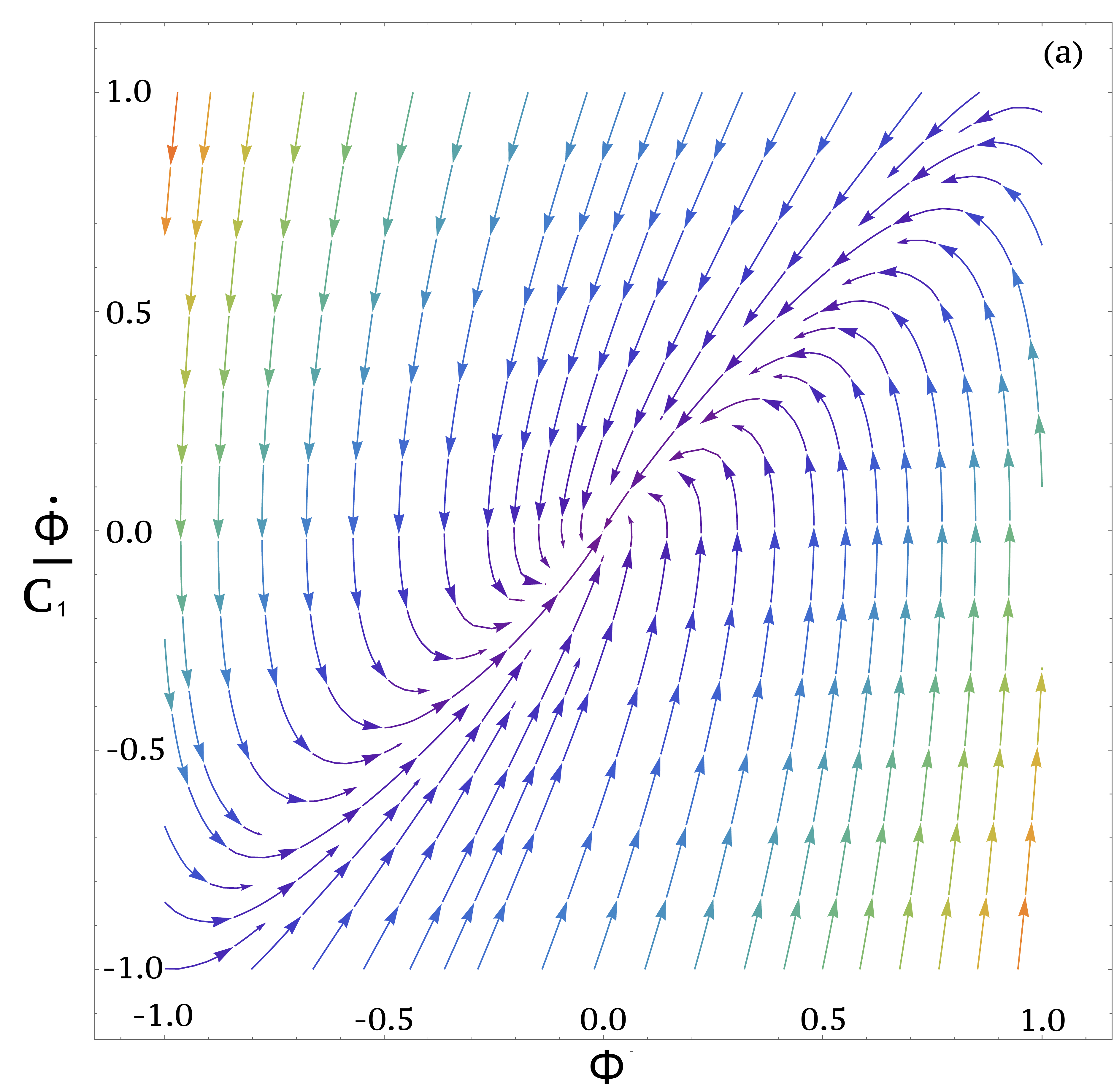}
\includegraphics[scale=.09]{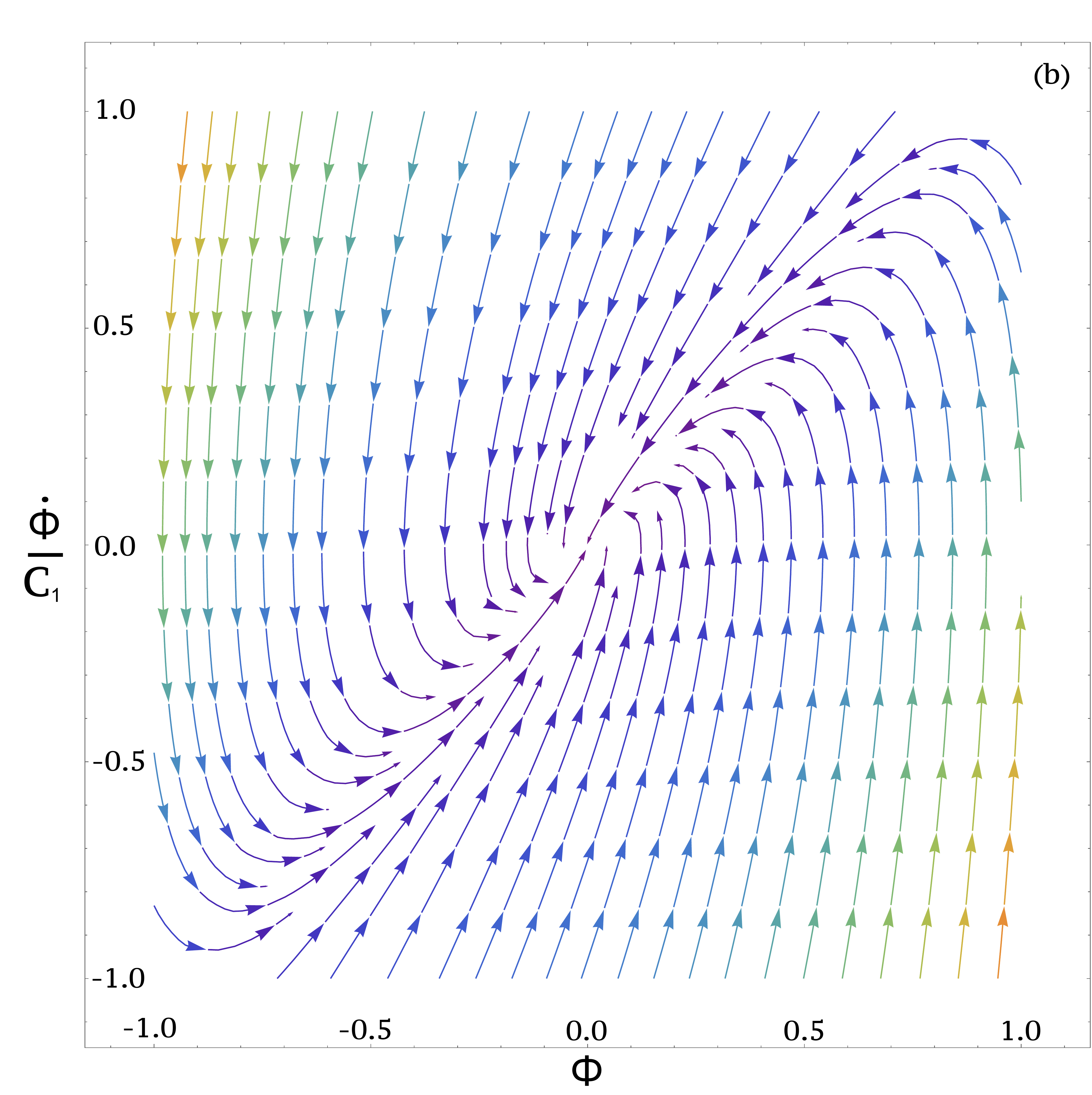}
\includegraphics[scale=.09]{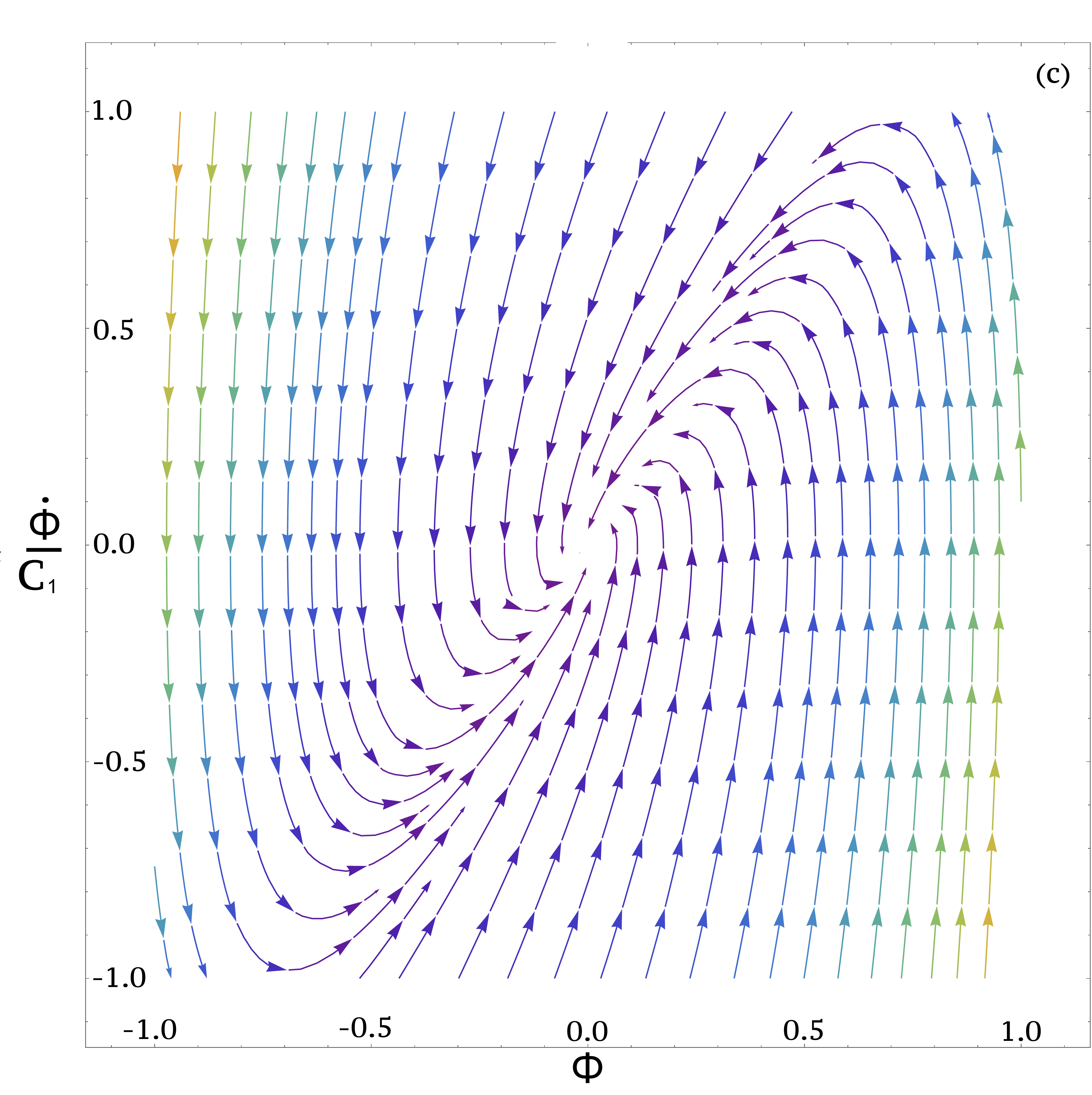}
\includegraphics[scale=.09]{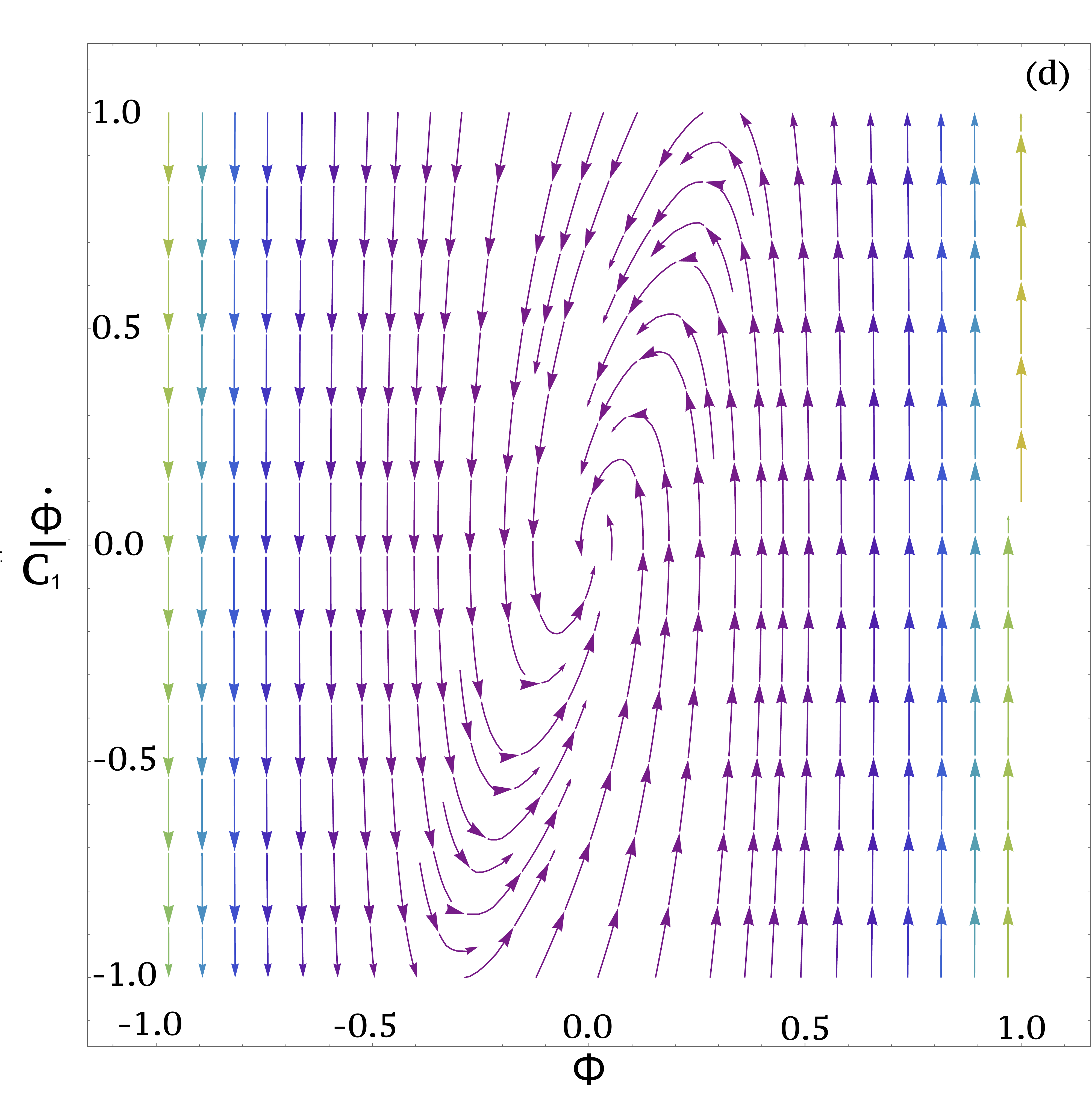}
\newline
\includegraphics[scale=.1]{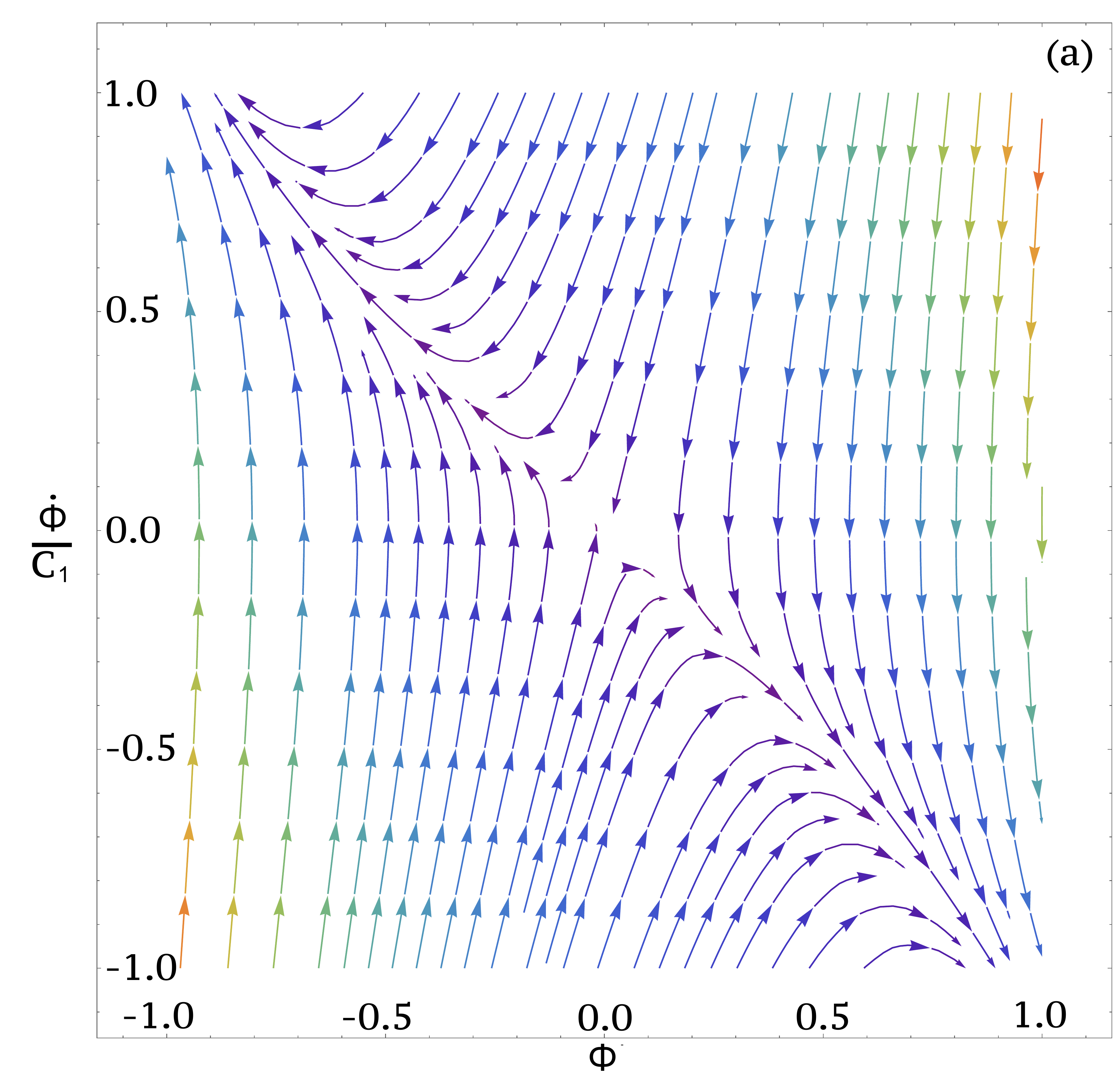}
\includegraphics[scale=.09]{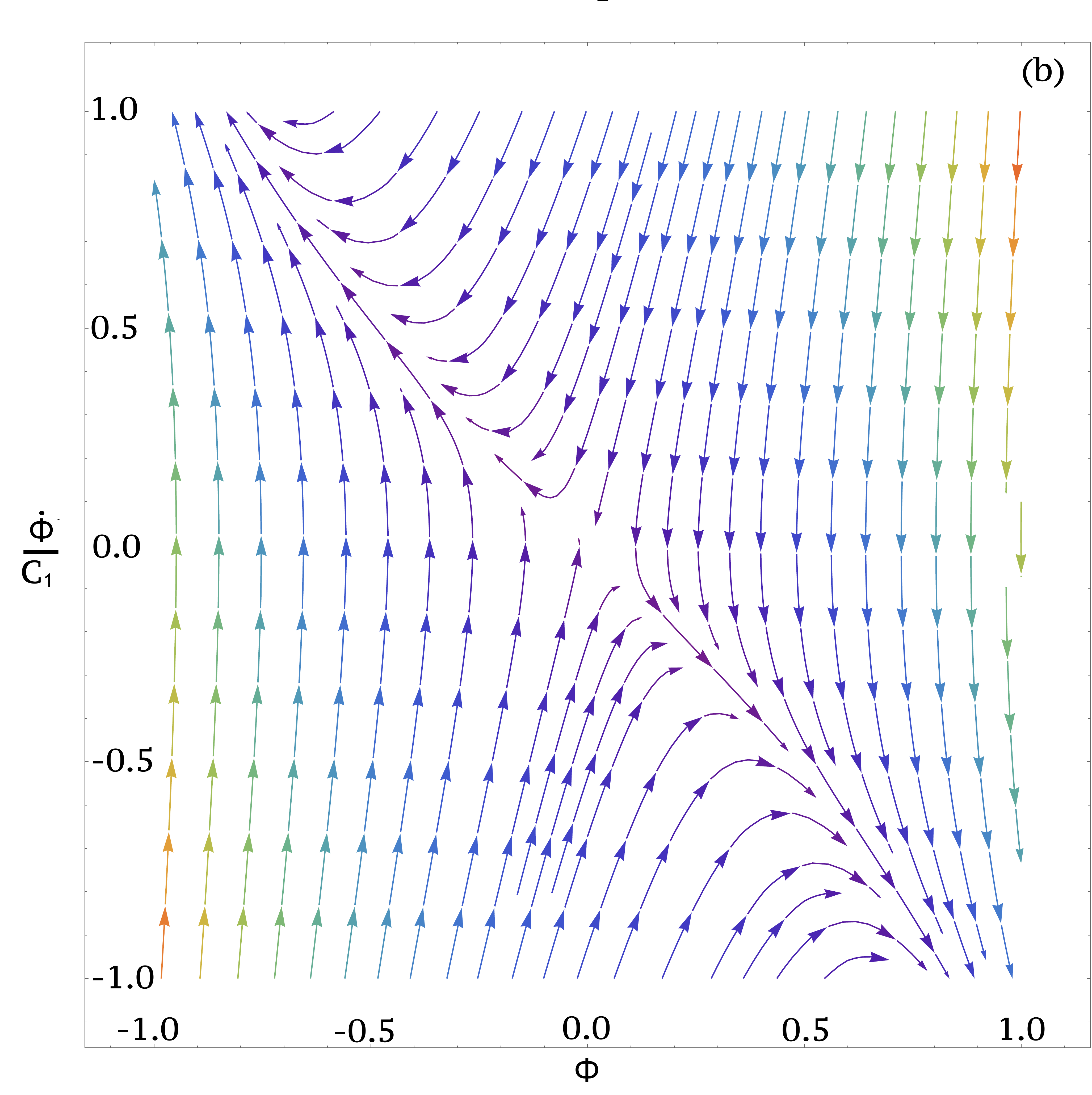}
\includegraphics[scale=.09]{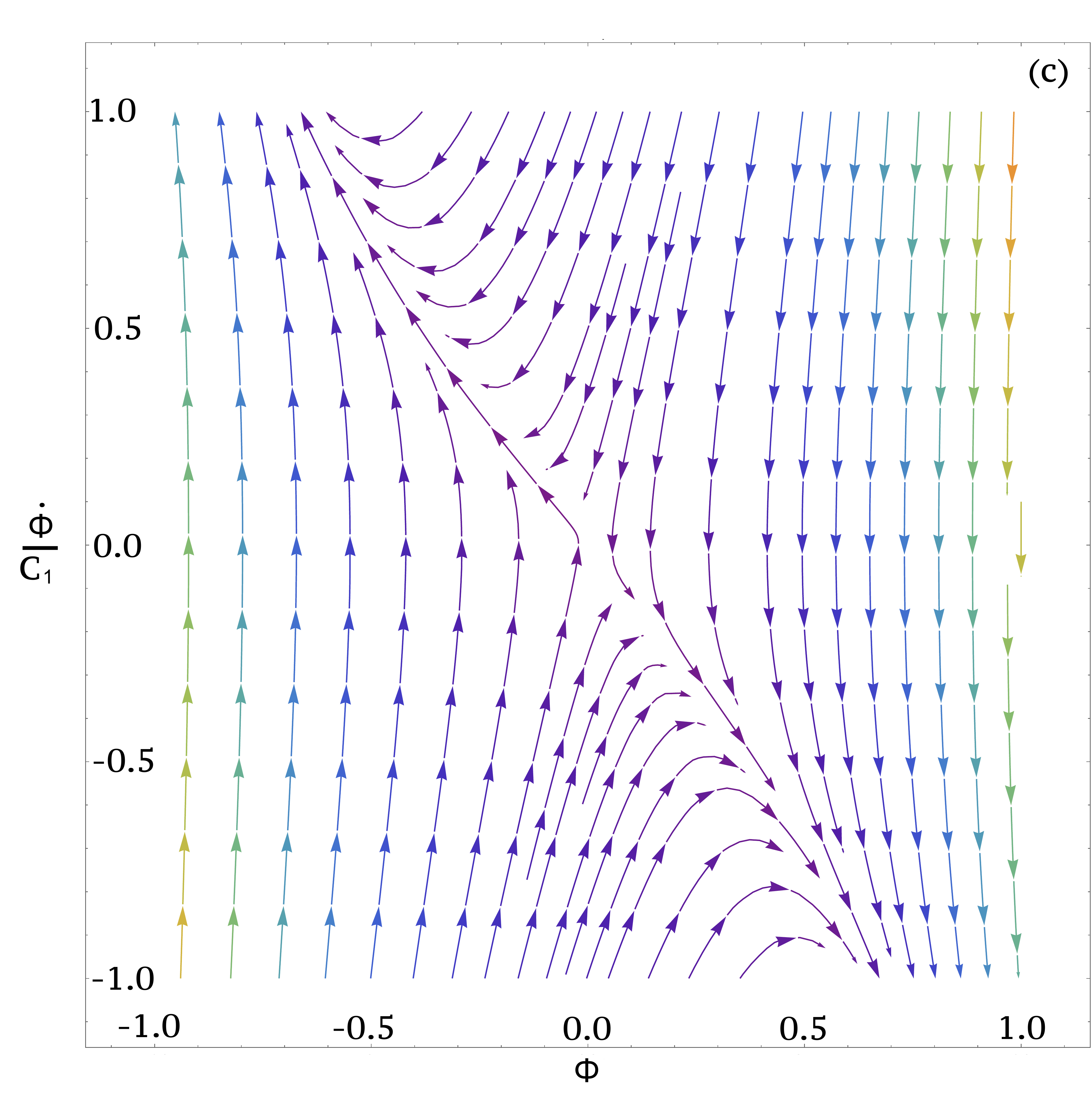}
\includegraphics[scale=.09]{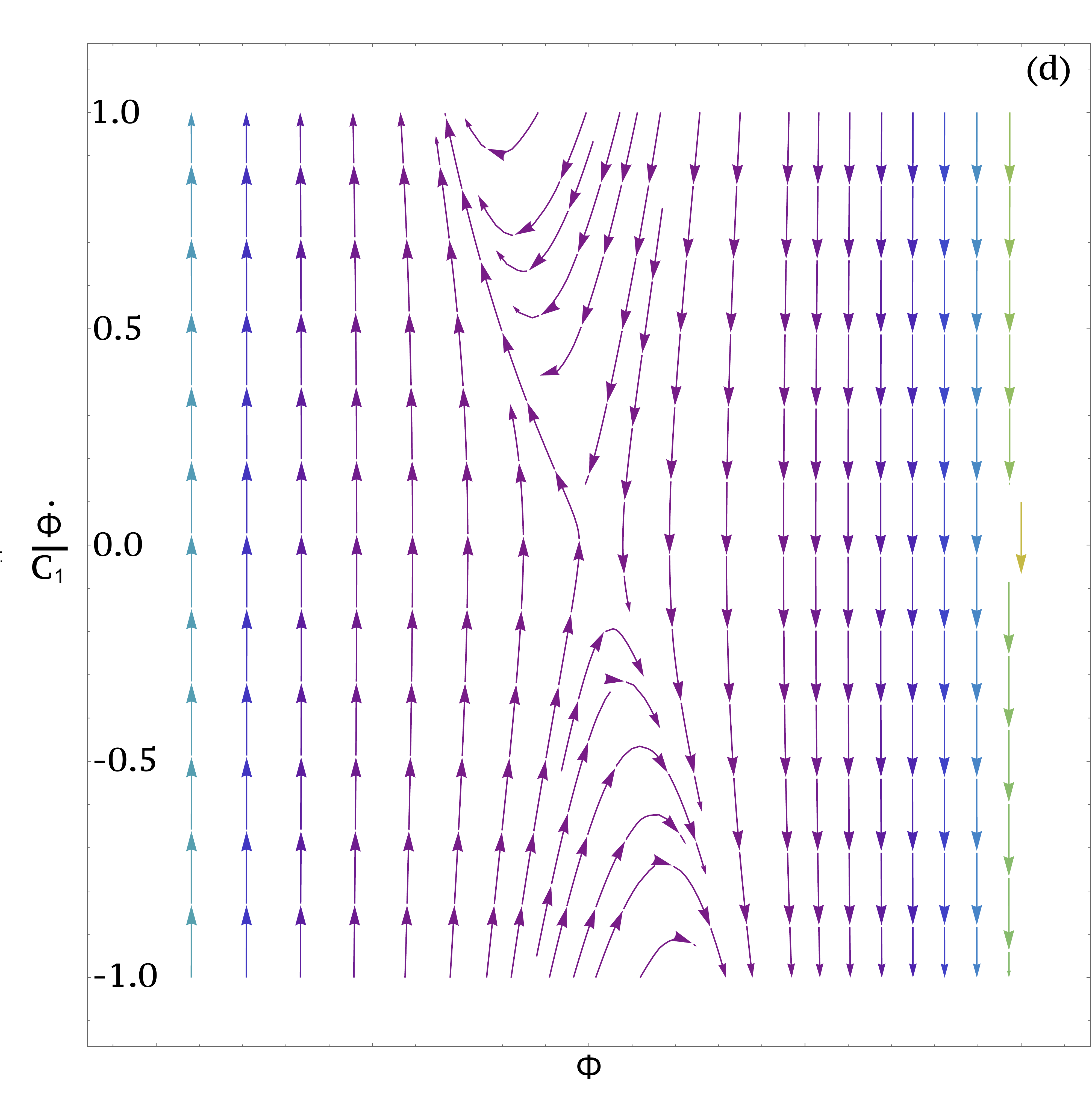}

\caption{Trajectories of Phase space determined by the analytical potential given in~\eqref{positive}. Plots in top and bottom rows are for $\eta=2.5,3.5$ respectively. Plots in the left to right columns are for $\gamma=0,\frac{\pi}{8},\frac{\pi}{4},\frac{\pi}{2.5}$ respectively. The domain of validity of our discussion is limited to the right upper half in these plots and the other parts are given for the comparison with $\gamma=0$.}
\label{fig:(gamma,eta)}
\end{figure}

The attractor solutions in phase space parametrized by $(y,x;\eta)$ should not give an indication of the stability of CR parameter $\eta$. Equation~\eqref{CR constraint 21} shows that this parameter is a measure of field acceleration and may well be time dependent in some phase space regions. Motivated by the duality relation $\eta\longleftrightarrow 3-\eta$ and the argument that slow roll is the unique attractor solution in all cases, the authors in~\cite{Lin:2019fcz} showed that for large $\eta$s, the constant roll solutions cannot be stable and the background perturbations will result in evolving $\eta$ to the smaller value of $\left\{\eta, 3-\eta\right\}$. In the following, we will look at the differently  parametrized phase space of CR complex fields to realize how the parameter $\gamma$ affects the asymptotic values of $\eta$. The stability of $\gamma$ under this stability analysis is assumed.

\begin{figure}
\includegraphics[width=\columnwidth]{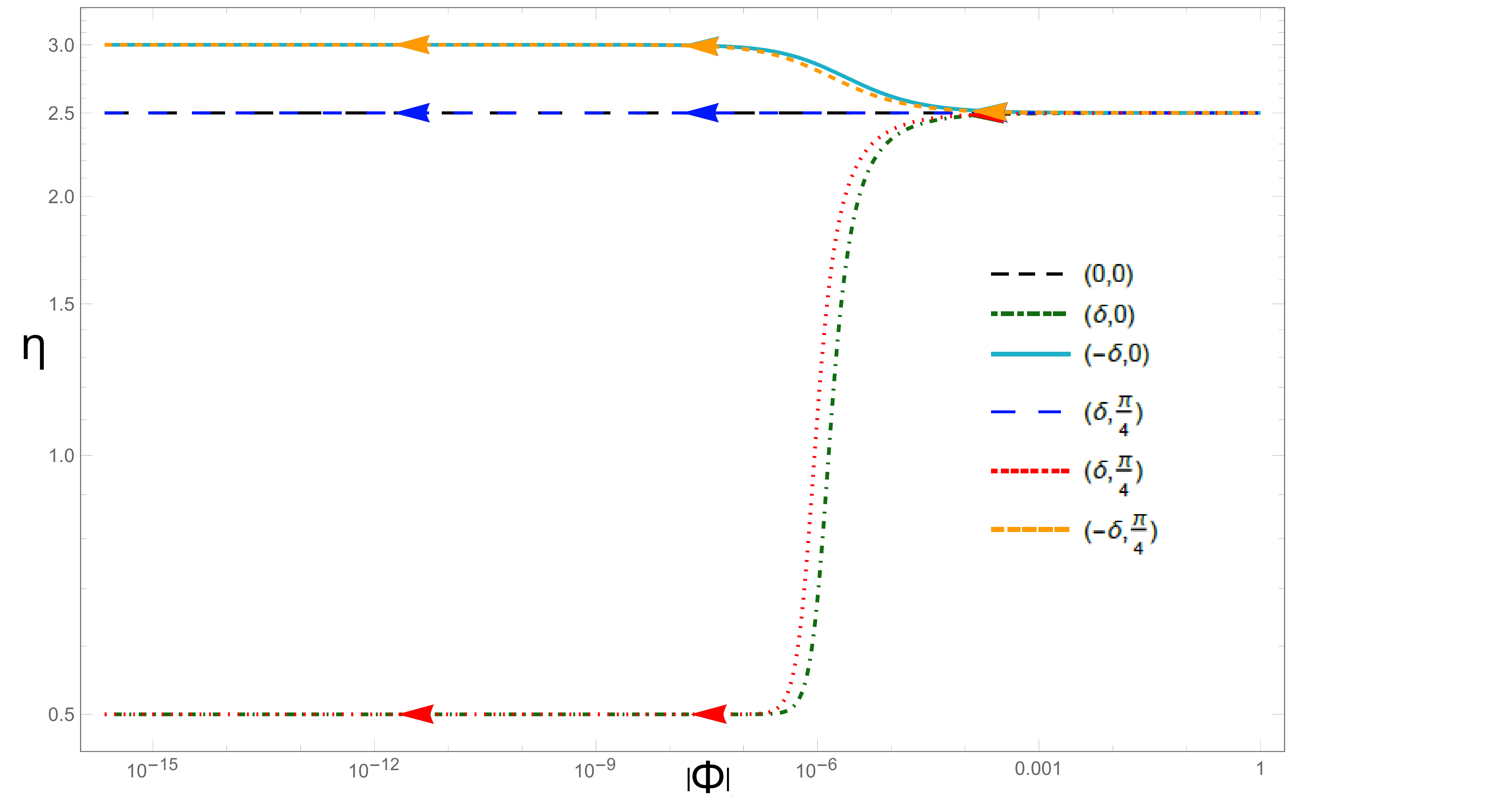}
\caption{Phase portrait of CR complex field parametrized by $\eta$, $|\Phi|$, for CR (convex) potential $\bar{\eta}=2.5$.}%
\label{phase space1}
\end{figure}

The phase space parametrized by $(y,x;\eta)$ breaks down at $V'\equiv{dV}/{dx}=0$ because the map $\eta(x,y)=3-\frac{V'(x)\cos(\gamma)}{H y}$ is not one-to-one at this point; so regardless of field velocity when the trajectories cross this point are forced to the critical value $\eta=3$. In other words, $\eta=3$ is the fixed point in $(x,\eta)$ plane. To check whether it is the late time attractor or not, the authors in~\cite{Pattison:2018bct} work on the variations of field acceleration around the fixed point. Following this methodology, we recast equation~\eqref{KG4} with $x$ and get the following first-order differential equation
\begin{equation}\label{dy/dx}
\frac{dy}{dx}-\frac{3H}{\cos(\gamma)}+\frac{V'}{y}=0,
\end{equation}
or
\begin{equation}\label{dy/dx equivalent}
\frac{dy^2}{dx}=-2V'\left(1-\frac{1}{\cos(\gamma)}\frac{3Hy}{V'}\right).
\end{equation}
Following the authors in \cite{Pattison:2018bct}, we also introduce the parameter $f=\frac{\eta}{3}$ and express equations~\eqref{dy/dx equivalent} and $y^2$ (or $|\dot{\Phi}|^2$) in terms of $x$ and $f$,
\begin{equation}\label{equations}
\frac{dy^2}{dx}=2V'\frac{f}{1-f},\qquad y^2=V\left[\sqrt{1+\frac{2}{3}\left(\frac{\cos(\gamma)}{1-f}\right)^2\left(\frac{V'}{V}\right)^2}-1\right].
\end{equation}
The combination of equations in~\eqref{equations}, leads us to an equation for the evolution of $f$,
\begin{equation}\label{df/dx}
\frac{df}{dx}=\frac{3}{2}\frac{V}{V'}\frac{(1-f)^2(1+f)}{\cos^2(\gamma)}\left[\sqrt{1+\frac{2}{3}\left(\frac{\cos(\gamma)}{1-f}\right)^2\left(\frac{V'}{V}\right)^2}-\frac{1-f}{1+f}\right]-(1-f)\frac{V''}{V'}.
\end{equation}
In order to study small deviations of complex fields from the $\eta=$ constant in phase plane, we linearise equation~\eqref{df/dx} around $f=\bar{\eta}/3$ by parametrizing
\begin{equation}
f=\frac{\bar{\eta}}{3}-\delta, \qquad |\delta|\ll1,
\end{equation}
and plot trajectories perturbed about a CR analytic solution started from $\bar{\eta}=2.5$ in figure~\ref{phase space1}.
\begin{figure}
\includegraphics[width=\columnwidth]{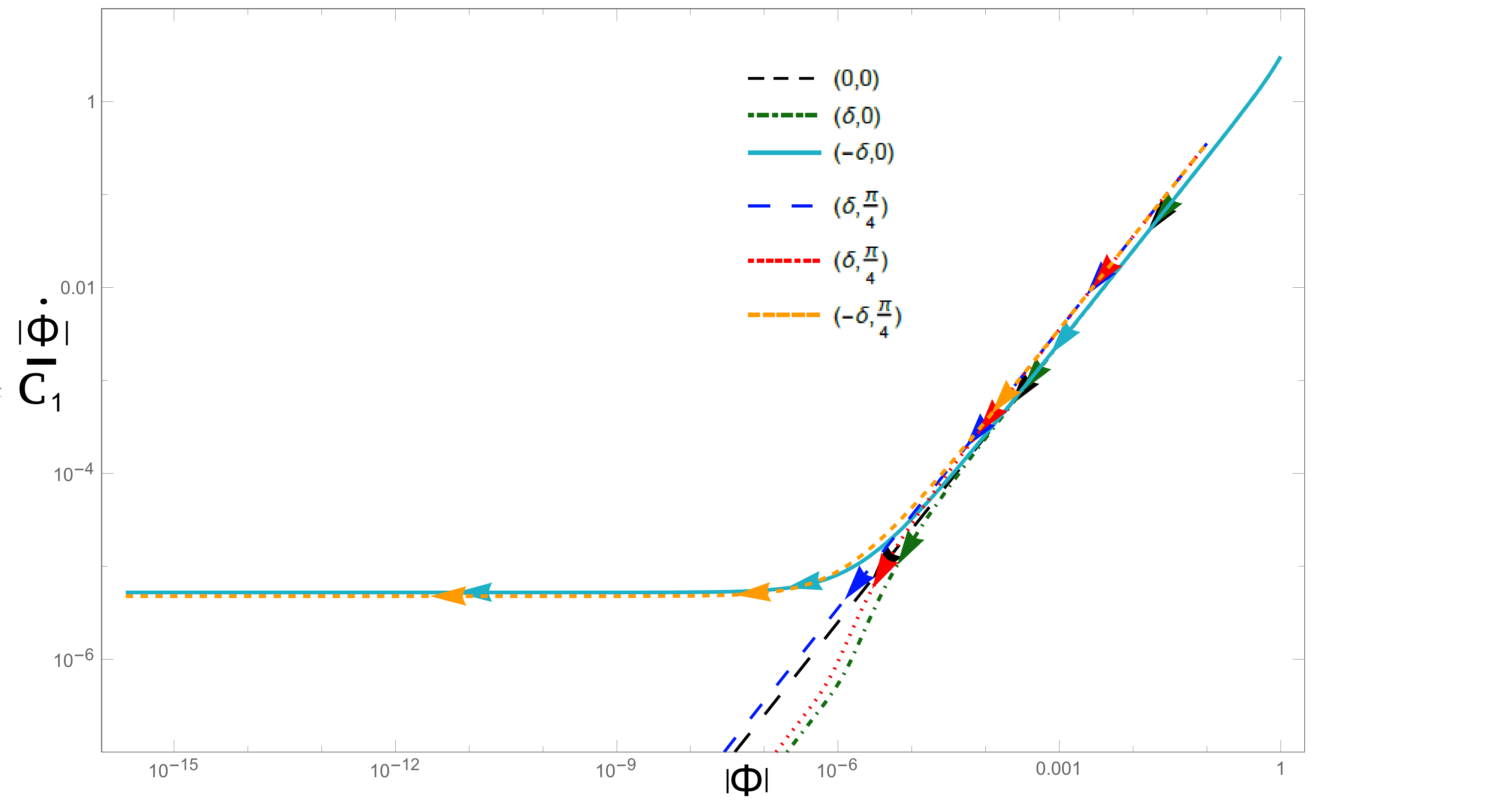}
\caption{Phase portrait of CR complex field parametrized by $|\dot{\Phi}|$, $|\Phi|$, for CR (convex) potential $\bar{\eta}=2.5$.}%
\label{phase space2}
\end{figure}
 In this plot we see one trajectory associated to the analytic solution that reaches the minimum point, at $|\Phi|=0$, with $|\dot{\Phi}|=0$\footnote{We will see that this trajectory with finely tuned initial condition is the only possible trajectory with constant roll as such.} and two other perturbed trajectories with deviations $\delta$ of order $10^{-5}$. For trajectories with $3f<2.5$ (or $\delta>0$), the field relaxes to slow-roll attractor with $\eta=0.5$, before reaching the minimum point. However, for $\delta<0$ enough field speed lets the field to pass the minimum point where $\eta=3$. The authors in~\cite{Gao:2019sbz} showed that if $\epsilon_1<\eta$, the real field rolls past this point and then turns back to it in slow roll. In complex field trajectories also $\eta$ evolves away from $2.5$ toward $0.5$, which is the stable solution as perceived by the duality.

The duality is valid at cosmological perturbation level as well.
 The evolution of the mode function $v_k=\sqrt{2}z\zeta_k$, with $z=a y/H$, is governed by the Mukhanov-Sasaki equation
\begin{equation}\label{Mukhanov-Sasaki}
v''_k+\left(k^2-\frac{z''}{z}\right)v_k=0,
\end{equation}
\begin{equation}
\frac{z''}{z}=a^2 H^2\left(2-\epsilon_1+\frac{3}{2}\epsilon_2+\frac{1}{4}\epsilon_2^2-\frac{1}{2}\epsilon_1\epsilon_2+\frac{1}{2}\epsilon_2\epsilon_3\right),
\end{equation}
and can be evaluated for constant roll solutions using
\begin{subequations}\begin{empheq}[left= \empheqlbrace]{align}
&\epsilon_1=2\kappa\tanh^2(\kappa x), \quad   & \epsilon_2=2\eta > 0, \qquad   & \epsilon_3=2\kappa\tanh(\kappa x),\label{positive1} \\
&\epsilon_1=-2\mu\tan^2(\mu x), \quad        & \epsilon_2=2\eta < 0, \qquad   & \epsilon_3=-2\mu \tan(\mu x)\label{negative1}
.\end{empheq}\end{subequations}
In either case we then have
\begin{equation}
\frac{z''}{z}=a^2 H^2\left[(\eta-2)(\eta-1)+\left(\eta^2-\frac{3}{2}\eta^3\right)\left(\frac{x}{\cos(\gamma)}\right)^2+O\left(\frac{x}{\cos(\gamma)}\right)^3\right].
\end{equation}
One can see the self-duality of Mukhanov-Sasaki equation under $\eta\rightarrow3-\eta$ in the limit of $\frac{|\Phi|}{\cos(\gamma)}\ll 1$. Outside this region the evolution of background and scalar perturbations are different and the duality between large and small $\eta$ breaks down. The authors in \cite{Gao:2019sbz} showed that the duality in scalar spectral tilt and tensor to scalar ratio expressions is true if $\epsilon_1<\eta<3$. Since the Hubble flow slow-roll parameters in CR complex and real field models are similar, it is easy to see that equations (12-15) in their work are valid for our model; so their result is valid for complex fields as well.
 
The additional DoF has just restricted the field value range of validity of the duality to a smaller region in the vicinity of potential minimum. This should come as no surprise, since the potential slope is increased and its linear approximation is restricted to the smaller region. It is interesting to see that there is nothing special about $\gamma=0$ in respecting the duality relation except that in this case complex fields leave the analytic (unperturbed) trajectory at smaller field values.

 For $\eta>3$, the analysis can be repeated as above. As $|\Phi|\rightarrow 0$, $|\delta|$ will grow causing $\eta$ to evolve away from the initial value either into slow roll (for $\delta>0$) or to $\eta=3$ (for $\delta<0$). In the former case the field rolls back the direction from which it came, down the hill and for the latter, the field rolls over the top of the potential. In both cases the generic attractor path is the dual slow roll regime with parameter $\bar{\eta}$ given by $3-\bar{\eta}$.
\subsection{Relative Kinetic Energy- Stability Analysis}\label{Stability Analysis}
The duality based  discussion in above subsection showed that slow roll is the unique attractor solution even in the presence of an additional DoF. In this subsection we study the stability of the solutions by evolving $\gamma$. We would like to know whether there are regimes that the phase of complex field has observational effects. Equation~\eqref{KG41} gives us the form of evolution of $\gamma$ as
\begin{equation}\label{gamma-dot}
\dot{\gamma}=-\frac{1}{\tan(\gamma(t))\frac{d|\Phi|}{dt}}\left[\frac{d^2|\Phi|}{dt^2}+\cos^2(\gamma(t))\frac{{dV}/{dt}}{{d|\Phi|}/{dt}}+3H(t)\frac{d|\Phi|}{dt}\right].
\end{equation}
The three terms in the square bracket that compete together in this evolution are: the acceleration of field, the force from the potential gradient and Hubble friction.

\begin{figure}
\includegraphics[scale=.59]{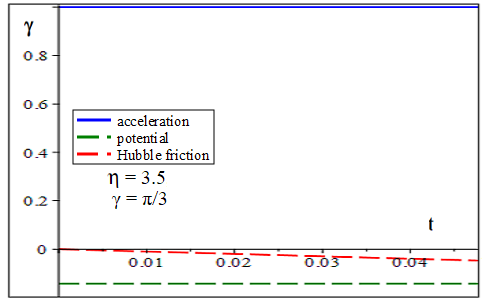}
\hspace{0.2cm}
\includegraphics[scale=.59]{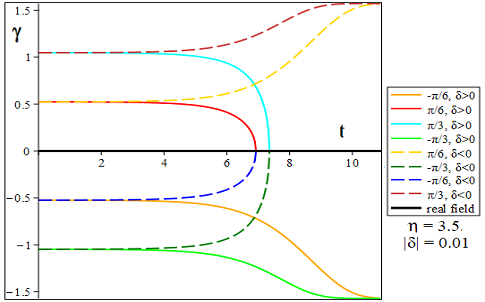}
\caption{A relative comparison between different terms in~\eqref{gamma-dot} for  $\eta=3.5$ is given in the left panel. The right panel shows the variation of $\gamma$ parameter and in particular the instability of $\eta$ = constant trajectories controlled by concave potentials. The first value in the legend corresponds to $\bar{\gamma}$ in each case.}\label{fig:gamma instability}
\end{figure}

Equation~\eqref{gamma-dot} is valid in the regime that the field has joined the attractor path on which $\eta$ = constant. We begin the stability analysis by substituting Hubble parameter ~\eqref{H(t)} for $\eta>0$ (or~\eqref{H1(t)} for $\eta<0$) and the corresponding potential given in ~\eqref{V(t)} into~\eqref{KG41}, but let the parameter $\gamma$ to be a function of time. We linearise equation~\eqref{gamma-dot} around a fixed value $\bar{\gamma}$ by
\begin{equation}
\gamma=\bar{\gamma}-\delta, \qquad |\delta|\ll 1
\end{equation}
and then plot trajectories in $\gamma-t$ space for different values of $\eta$. The results for different values of $\bar{\gamma}$ and $\eta=3.5$ are shown in figures \ref{fig:gamma instability}.
In this case, the trajectories with $\gamma$ = constant are unstable. The paths with $\delta<0$ asymptote to $\gamma=\pi/2$ (or $\gamma=-\pi/2$) and for $\delta>0$, the paths ends up to $\gamma=0$. The contribution of different terms of square bracket in~\eqref{gamma-dot} is also shown in figure \ref{fig:gamma instability}. For $\eta>3$, the force driven by the concave potential and the frictional force in~\eqref{gamma-dot} have equal signs; so any small change in $\gamma$ (or energy proportions) will be amplified to reach the asymptotic value. 

\begin{figure}
\includegraphics[scale=.59]{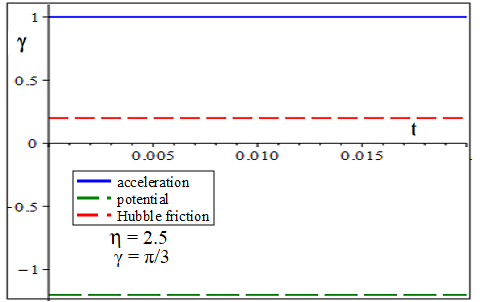}
\hspace{0.2cm}
\includegraphics[scale=.59]{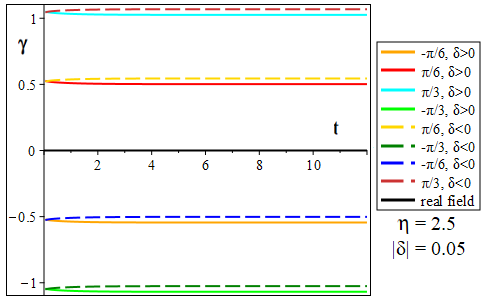}
\newline
\includegraphics[scale=.59]{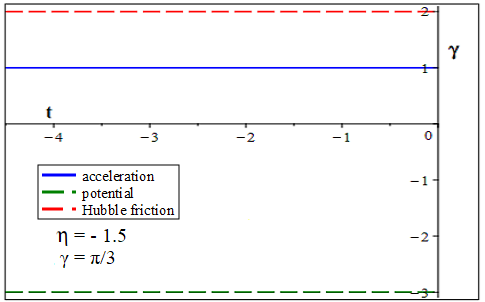}
\hspace{0.2cm}
\includegraphics[scale=.59]{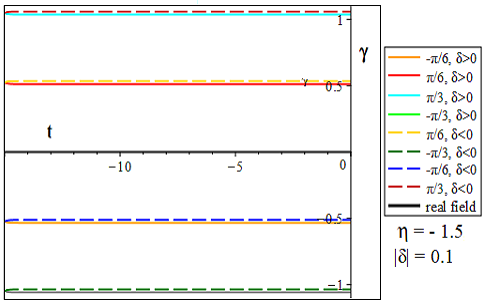}
\caption{A relative comparison between different terms in~\eqref{gamma-dot} for values of $\eta=2.5$ and $-1.5$ is given in the left column. The right column shows the stability of $\eta$ = constant trajectories directed by convex potentials for these values of $\eta$. The first value in legends corresponds to $\bar{\gamma}$ in each case.}\label{fig:gamma stability}
\end{figure}

The situation is completely different when the convex potentials play role. As it is shown in figure \ref{fig:gamma stability}, for $\eta<3$, the trajectories with $\gamma$ = constant are stable. On these trajectories the three above mentioned terms cancel each other and any small changes in energy proportion will soon freeze. Figures \ref{fig:gamma instability} and \ref{fig:gamma stability} show that the (in)stability is controlled by the sign and magnitude of potential gradient which is invariant under $\bar{\gamma}\leftrightarrow -\bar{\gamma}$. The symmetry is seen in the (in)stability behavior of system solutions.

Naively, then we must conclude from this surprising result that the concave potential given in~\eqref{positive} should be rectified for $\eta>3$. However, this result is only valid on the attractor path set by the duality relation $\eta\longrightarrow\left\{\eta,3-\eta\right\}$. From the analysis about the cosmological perturbations given in \cite{Gao:2019sbz}, it is easy to see that the time profile of mode functions $v_k$, the spectral tilt and the tensor to scalar ratio depend on the Hubble flow slow-roll parameters, which is similar to that of a real CR field and is stable under $\gamma$ variations.
 We, therefore conclude that for the field values far from the range of validity of the duality, $\gamma$ = constant are stable and the potential given in~\eqref{eqn:V(Phi)} very well describe the CR complex field inflationary models.

\section{General Multifield Constant Roll Inflation}\label{multifield}

In previous sections, we showed under what conditions a-priori overdetermined system of equations in CR inflation with complex scalar field can be consistent. In this section we follow the similar approach in the study of general multifield CR Inflation. For the geometrical description of general field space in multifield inflation, we follow the covariant formalism discussed in \cite{GrootNibbelink:2000vx,GrootNibbelink:2001qt,Gong:2011uw,Pinol:2020kvw}. We consider the following four dimensional action  consisting of gravity and a set of $N$ minimally coupled real scalar fields, $\phi^I$ $\left(I=1,...,N\right)$
\begin{equation}
S=\int{\sqrt{-g}d^4 x\left(\frac{R}{2}-\frac{1}{2}g^{\mu\nu}G_{IJ}\left(\partial_\mu \phi^I\right)(\partial_\nu \phi^J)-V\left(\phi^I\right)\right)}.
\end{equation} 
These fields span a scalar manifold $\mathcal{M}$ of dimension $N$, equipped with a scalar metric $G_{IJ}$. The geometry of spacetime given by spatially flat FLRW. The Friedmann equations are given by
\begin{equation}\label{Friedmann N-field}
3H^2=\frac{1}{2}\dot{\sigma}^2+V(\phi^I), \qquad -2\dot{H}=\dot{\sigma}^2.
\end{equation} 
 We can think of a background field trajectory parametrized by a single parameter, usually taken as the cosmic time $t$, i.e, $\phi^I=\phi^I(t)$. Here, $\dot{\sigma}^2\equiv G_{IJ}\dot{\phi}^I \dot{\phi}^I$ is the total squared velocity in the field space and is related to $Z$ defined in \eqref{Z definition} by $Z=\dot{\sigma}^2$. In this background, the equations of motion governing the homogeneous scalar fields are 
\begin{equation}\label{equation of motion-N fields}
\frac{\mathcal{D}}{dt}\dot{\phi}^I+3H\dot{\phi}^I+V^{,I}=0,  
\end{equation}
where $V^{,I}=G^{IJ}\frac{\partial V}{\partial \phi^J}$ and $\frac{\mathcal{D}}{dt}\dot{\phi}^I=\ddot{\phi}^I+\Gamma^I_{JK}\dot{\phi}^J\dot{\phi}^K$ is the covariant derivative with respect to spacetime coordinate $t$ on the vector $\dot{\phi}^I$. Once the scalar potential $V$ and the metric $G_{IJ}$ are specified, the classical trajectory in the background can be determined. To study this trajectory, it is convenient to work with the vielbeins $\left\{\textbf{e}_n\right\}$, on $\mathcal{M}$, first introduced in \cite{GrootNibbelink:2000vx}. The first unit vector $\textbf{e}_1$, which is denoted by adiabatic direction, is tangent to the trajectory and we have $\dot{\phi}^I=\dot{\sigma}e_1^I$. The rest of the unit vectors $\textbf{e}_n$, called the entropic sector, span the part of field acceleration $\frac{\mathcal{D}}{dt}\dot{\phi}$ that is orthogonal to $\textbf{e}_1$. We use lowercase letters $a,b=2...N$ for entropic indices and decompose the metric on the field space by $G^{IJ}=\delta^{ab}e^I_a e^J_b=e^I_1 e^J_1+\sum_{a=2}^{N}e^I_a e^J_a$. Such local orthonormal basis is not unique and in fact any orthogonal rotation preserves these properties; so there is an ambiguity in defining the directions perpendicular to the instantaneous direction of the background. To fix this ambiguity, the first entropic direction is defined to be along the $\frac{\mathcal{D}}{dt}\dot{\phi}$ and all other vectors in this sector can be defined as the wedge product of the two previous vectors. This results in the following form for the covariant rate of turn of these vielbeins;
\begin{subequations}\begin{align}
\frac{\mathcal{D}}{dt}\textbf{e}_1&=\omega_1 \textbf{e}_2,\\
\frac{\mathcal{D}}{dt}\textbf{e}_a&=-\delta_{1a}\omega_1 \textbf{e}_1+\Omega_{a}^{\ b}\textbf{e}_b   \qquad  a,b=2...N.
\end{align}\end{subequations}
$\Omega$ is an anti-symmetric rotation matrix of size $(N-1)\times(N-1)$. The non-zero components of $\Omega$ and $\omega_1$ are called "`mixing parameters" and have the dimension of mass. Those can be interpreted as the turning rate of local basis which guarantees that the relations $G^{IJ}e^a_Ie^b_J=\delta^{ab}$ and $\delta^{ab}e^I_a e^J_b=G^{IJ}$ be valid at any time along the background trajectory. The projection of Equation \eqref{equation of motion-N fields} along these orthogonal directions gives us \cite{Pinol:2020kvw}
\begin{subequations}\begin{align}
 \ddot{\sigma}+3H\dot{\sigma}+V_\sigma=0,\label {sigma dynamics}\\   \qquad  \dot{\sigma}\omega_1\delta_{a,2}=-V_a \label{normal dynamics-general}.
\end{align}\end{subequations}\label{}
Here $V_\sigma=e_1^I V_{,I}$ and $V_a=e_a^I V_{,I}$ are the components of potential gradients along the adiabatic and entropic directions and the mixing parameters are the generalized curvature of the trajectory.  With this choice of entropic sector, $V_a$ is non-zero only in the direction of $\textbf{e}_2$.

For two-field case, $N=2$, unit vectors $T^I\equiv\textbf{e}_1$ and $N^I\equiv\textbf{e}_2 $ are defined as
\begin{equation}
 T^I=\frac{\dot{\phi}^I}{\dot{\sigma}},   \qquad   N^I=\sqrt{det G}\epsilon_{JK} G^{IJ} T^K.
\end{equation}

Now, focusing on the time (covariant) derivative of $T^I$, we have 
\begin{equation}\label{covariant time derivative}
\frac{\mathcal{D}}{dt}T^I=\frac{1}{\dot{\sigma}}\left(\frac{\mathcal{D}}{dt}\dot{\phi}^I\right)-\frac{\ddot{\sigma}}{\dot{\sigma}^2}\dot{\phi}^I=-\left(3H+\frac{\ddot{\sigma}}{\dot{\sigma}}+\frac{V_\sigma}{\dot{\sigma}}\right)T^I-\frac{V_N}{\dot{\sigma}}N^I.
\end{equation}
Here, $V_\sigma\equiv T^I V_{,I}$ and  $V_N\equiv N^I V_{,I}$ correspond to the components of $V_{,I}$ in $T^I$ and $N^I$ directions, respectively and we have used the equation of motion \eqref{equation of motion-N fields} in the second equality. The projection of Equation \eqref{covariant time derivative} along these orthogonal directions gives us
\begin{subequations}\begin{align}
 \ddot{\sigma}+3H\dot{\sigma}+V_\sigma=0,\label {sigma dynamics}\\   \qquad  \frac{\mathcal{D}}{dt}T^I=-\frac{V_N}{\dot{\sigma}}N^I \label {normal dynamics}.
\end{align}\end{subequations}\label{equations of motion-multified-field}
These two independent equations well characterize the dynamics of the two-field trajectory. These also show a separation between a single $\it{effective}$ dynamical field $\sigma$\footnote{The field $\sigma$ is defined as a curvilinear coordinate in $\mathcal{M}$ by the time integral of $\dot{\sigma}$.} and truly multiple fields contributions in field dynamics. 
 In our approach to CR inflation, we impose the CR constraint on the effective field dynamics and the equation \eqref{normal dynamics} serves us as the consistency condition\footnote{The CR constraint affects solely on the tangent component of $V^{,I}$.}. The latter measures the deviation of the background trajectory from a geodesic in the field space, $\mathcal{M}$. If $V^{,I}=V_\sigma T^I$, then $\frac{\mathcal{D}}{dt}T^I=0$ and the trajectory will coincide with such geodesic and $\left\{T^I,N^I\right\}$ remain covariantly unchanged along the trajectory. Otherwise the changes in $T^I$ will be balanced by $V_N$. 

Let us find the consistency relation in a \textit{complex field} model following the general approach we introduced in this section. In this model, which is an example of $N=2$ case with $I=(X,\theta)$ and $G_{IJ}=$diag$\left[1,X^2\right]$, we have $\dot{\sigma}^2=\dot{X}^2+X^2\dot{\theta}^2$ and the only non-zero Christoffel symbols are $\Gamma^X_{\theta\theta}=-X$ and $\Gamma^\theta_{X\theta}=\Gamma^\theta_{\theta X}=\frac{1}{X}$. The tangent and normal vectors to the field trajectory are given by 
\begin{equation}
T^I=\frac{1}{\dot{\sigma}}(\dot{X},\dot{\theta}),  \qquad N^I=\frac{X}{\dot{\sigma}}(\dot{\theta},-\frac{\dot{X}}{X^2}).
\end{equation}
In subsection \ref{sec: consistency condition}, we used $X\dot{\theta}=\lambda\dot{X}$, which is equivalent to $\dot{X}=\dot{\sigma}\cos\gamma$, $X\dot{\theta}=\dot{\sigma}\sin\gamma$ (with $\lambda=\tan \gamma$) in non-canonical field basis.  For $I=X$, we have
\begin{subequations}\begin{align}
\frac{\mathcal{D}}{dt}T^X&=-\frac{\ddot{\sigma}}{\dot{\sigma}^2}\dot{X}+\frac{\ddot{X}}{\dot{\sigma}}-\frac{X\dot{\theta}^2}{\dot{\sigma}}=-\frac{\lambda^2\dot{X}^2}{X\dot{\sigma}},\label{LHS}\\
V_N N^X&=\frac{\lambda\dot{X}^2}{X\dot{\sigma}^2}\left(\lambda X\frac{\partial V}{\partial X}-\frac{\partial V}{\partial \theta}\right)\label{RHS}.
\end{align}\end{subequations}
In the second equality of~\eqref{LHS}, we used $\ddot{\sigma}/\dot{\sigma}=\ddot{X}/\dot{X}$. Following~\eqref{normal dynamics}, it is easy to see that the consistency relation is given by $\left(\lambda X\frac{\partial V}{\partial X}-\frac{\partial V}{\partial \theta}\right)=\lambda\dot{\sigma}^2$, which is the same as~\eqref{consistency condition}. For single scalar field models, $\lambda=0$ and we have $\frac{\mathcal{D}}{dt}T^I=0$. No consistency relation is required in these models and equation~\eqref{normal dynamics} trivially holds. 

 Using the CR constraint~\eqref{CR constraint4} written in the form of $\ddot{\sigma}=-\eta H\dot{\sigma}$, we find that
\begin{equation}
\frac{\mathcal{D}}{dt}\dot{\phi}^{I}=-\eta H\dot{\phi}^{I}-\frac{1}{2\dot{\sigma}^2}G^{KI}(\frac{\mathcal{D}}{dt}G_{KJ})\dot{\phi}^{J}.
\end{equation}
  
If the potential is \textit{assumed} to have a sum separable form, $V=\sum_{I=1}^N V_I(\phi^{I})$, we will have $\partial V/\partial \phi^I=dV_I/d\phi^I$ and equations of motion~\eqref{equation of motion-N fields} give us
\begin{equation}\label{potential- multifield}
V_I(\phi^{I})=\int_{\mathcal{M}}\frac{dV}{d\phi^{I}}d\phi^{I}=\int_{\mathcal{M}} \left(\eta-3\right)H G_{IJ}\dot{\phi}^J d\phi^{I}+\frac{1}{2}\int_{\mathcal{M}}\dot{\phi}^J(\frac{\mathcal{D}}{dt}G_{IJ})d\phi^{I}.
 \end{equation}
It is easy to recognize the different component of $V_{,I}$ in this potential form. The integrand in the first integral of~\eqref{potential- multifield} is parallel to $\textbf{e}_1$ and corresponds to $V_\sigma$, whereas the component in entropic direction is given by $V_a=\frac{1}{2\dot{\sigma}^2}(\frac{\mathcal{D}}{dt}G_{IJ})\dot{\phi}^J e^I_a$. The latter vanishes if we have $\frac{\mathcal{D}}{dt}G_{IJ}=kG_{IJ}$.

 It is possible to show that equation~\eqref{normal dynamics-general} is, surprisingly, \textit{always} satisfied by this potential form. Using the definition $\dot{\phi}^I=\dot{\sigma}e_1^I$, the covariant rate of change of $\dot{\phi}^I$ is given by $\frac{\mathcal{D}}{dt}\dot{\phi}^{I}=-\eta H e_1^{I}+\dot{\sigma}\omega_1e^I_2$. The sum separable form of the potential allows us to write 
\begin{subequations}\begin{align}
V_{,2}=V_{,I}e^I_2&=G_{IJ}\left(-3H\dot{\sigma}e^J_1-\frac{\mathcal{D}}{dt}\dot{\phi}^{J}\right)e^I_2\nonumber\\
&=G_{IJ}\left((\eta-3\dot{\sigma})He^J_1-\dot{\sigma}\omega_1e^I_2\right)e^J_2\nonumber\\
&=-\omega_1\dot{\sigma}
\end{align}\end{subequations}
In the last equality, we used the orthonormal property of vielbeins. To summarize our findings, the system of evolution equations in a multifield CR model is inherently overdetermined. We, however showed that with the special form of potential the entropic sector of theory is dynamically decoupled from the effective field. The resulting single field theory is practically unaffected by multifield effects.

In the following, Let us make more precise a procedure whereby the entropic sector can be decoupled. If our assumption about the form of the potential is valid, there should be a relation between each pair of $\phi^{I}$s and it is necessary to know the Hubble parameter as a function of $\phi^{I}$. The time dependence of Hubble parameter is given by~\eqref{H(t)}, but the point is that its field dependence is ambiguous because it is not clear how to find $t(\phi^{I})$\footnote{Motivated by the string/supergravity models, the authors in \cite{Micu:2019fju} studied the CR behavior of a system comprising two scalar fields with a non-trivial metric on the field space. The CR constraint imposed on the system of equations is identical to the one we use in this paper. However, the way we face up to the ambiguity in $t(\phi^{I})$ is different from what is carried out in that article. In \cite{Micu:2019fju}, the potential function dependency to different fields is controlled by a function $f(t)$. This function is found by imposing the ansatz $V=(1-f(t))V(\phi^1)+f(t)V(\phi^2)$ on one of equations of motion. In this approach, the time parameter enters the potential function and the prescription of how to choose $t$ as a specific function of $\phi^1$ and $\phi^2$ is not clear.}.
 In the complex field CR inflation model studied in previous sections, an axillary assumption on the canonical fields helped us to find $t(\phi^{I};\gamma)$ and $\phi^{I}=(\phi^{J};\gamma)$.
To do so in general, we use another set of vielbeins $e^I_{(\alpha)}$ of the field space, called parallel transported vielbeins, and therefore $\frac{\mathcal{D}}{dt}e^I_{(\alpha)}=0$. In this local orthogonal frame, we have 
\begin{equation}
\delta^{\alpha\beta}e^I_{(\alpha)} e^J_{(\beta)}=G^{IJ},    \qquad   G_{IJ}e^I_{(\alpha)} e^J_{(\beta)}=\delta_{(\alpha)(\beta)}.
\end{equation}
We suppose that the contribution of different canonical fields in $\dot{\sigma}^2$ are given by
\begin{equation}\label{equal partitioning}
{\displaystyle {\begin{aligned}\dot{\phi}^{(1)}&=\dot{\sigma}\cos(\gamma _{1})\\\dot{\phi}^{(2)}&=\dot{\sigma}\sin(\gamma _{1})\cos(\gamma_{2})\\\dot{\phi}^{(3)}&=\dot{\sigma}\sin(\gamma_{1})\sin(\gamma _{2})\cos(\gamma _{3})\\&\,\,\,\vdots \\\dot{\phi}^{(N-1)}&=\dot{\sigma}\sin(\gamma _{1})\cdots \sin(\gamma _{N-2})\cos(\gamma _{N-1})\\\dot{\phi}^{(N)}&=\dot{\sigma}\sin(\gamma _{1})\cdots \sin(\gamma_{N-2})\sin(\gamma_{N-1}).\end{aligned}}}
 \end{equation}
in which $N-1$ arbitrary functions $\gamma_i$, $i=1,...,N-1$ are the natural generalization of the function $\gamma$ introduced in Section \ref{CR Potential}. 
  As before, we presume that the parameters are constants. This would help in resolving the ambiguity: By using the time profile of $\dot{\sigma}$, given in ~\eqref{Z(t)}, $\phi^{(\alpha)}(t;\vec{\gamma})$ and then $t(\phi^{(\alpha)};\vec{\gamma})$ are found. In the next step, $\dot{\phi}^{(\alpha)}(\phi^{(\alpha)})$, $H(\phi^{(\alpha)})$ and then $\dot{\phi}^J(\phi^I)$ and $H(\phi^I)$ can be determined for each field $\phi^I$. With those at hand, the integrand in~\eqref{potential- multifield} is given in terms of $\varphi^I$ and free parameters and the potential function can be found. This potential has $N$ parameters denoted by $(\eta,\gamma_i)$, but the key point is that the potential in this form is compatible with the constant roll condition~\eqref{CR constraint3}, the single effective field dynamics as well as the consistency relations (normal dynamics).
 
Before closing this section, we would like to give a justification for the free parameters in the potential. The curvature of field space implies the existence of couplings in kinetic terms of each field and the free parameters control the couplings of each field in the inflation dynamics. To discern the role of free parameters, we find the two-field potential for canonical fields, in Appendix~\ref{sec:canonical}. $\dot{\sigma}^2$ is a positively homogenous function of second degree, when expressed in canonical coordinates. We, therefore, expect to see the rotational symmetry (hence equal contribution of fields) in the potential, $V=3H^2+\dot{H}$, as well. There we show that the CR potential has only one parameter, $\eta$. Following the same steps the generalization of this potential to beyond two-fields model is straightforward 
\begin{equation}\label{potential-multifield}
V({\vec{\phi}})=(3-\eta)\left[C_1^2 e^{\sqrt{2\eta\sum_{\alpha=1}^N \phi_{(\alpha)}^2}}+C_2^2 e^{-\sqrt{2\eta\sum_{\alpha=1}^N \phi_{(\alpha)}^2}}\right]+2C_1 C_2(3+\eta).
\end{equation}

The CR inflation with two fields and trivial metric $G_{IJ}=\delta_{IJ}$, has been studied in \cite{Guerrero:2020lng}. The authors analyzed two cases: case 1) the CR parameters on the two fields are the same and case 2) where those are different. The first case is similar to our discussion in the appendix~\ref{sec:canonical}. For the second case, however, apart from the unclear motivation, the potential and Hubble parameter functions are given as functions of the number of e-folding and no prescription for $V(\vec{\varphi})$ is given. Furthermore, in the conclusion, the authors clearly state that” the differences between the constant roll parameters of both fields do not lead to an adiabatic field that also constant rolls”. We note that this result is not in contrast with multifield CR models discussed above, because this case cannot be described by the constraint~\eqref{CR constraint3}. We stress that our general approach in this article for determining the CR potential can be used to any number of fields and field space metric constrained by~\eqref{CR constraint3}.

\section{Summary and Conclusion}\label{Conclusion}

In a $N$-field cosmological scenario, the dynamical system includes $N+1$ equations governing $N$ matter fields and background scale factor. In this work we discussed that, for $N>1$, solving the coupled non-linear equations for the scalar and gravitational fields needs more caution. 
Imposing a relation between value and velocity of fields may result in an over-determined system with generally no solution. In other words, not every constraint, that seems to be physically relevant, is compatible with the $N+1$ dynamical equations and renders the system integrable. In a subset of non-slow roll inflationary models known as constant roll inflation, we worked on complex field dynamical $\Phi$ equations and found a class of models for an appropriate constant roll definition. In these models there exists a degree of flexibility in the form of a function of time, $\gamma(t)$, which makes it possible to define an isomorphism between the field DoFs at each instant of time.

Apart from the mathematical difficulties associated with solving dynamical equations with a general $\gamma(t)$, choosing such a function should be based on physical grounds. In an attempt to surmount the mathematical difficulties, we first focused on a subclass of solutions with an arbitrary but constant value of parameter $\gamma$, where $\gamma=0$ corresponds to the real field model. In this subclass, the $V(\Phi;\gamma)$ potential functions have larger curvature, if compared to the real field CR models at a fixed $\Phi$ value. By performing the stability analysis on the large $\eta$ phase space solutions, we showed that, similar to a real field, this subclass solutions are at most early time transients. Small perturbations to these solutions deviate the phase space trajectory with CR parameter $\eta$ from the larger unstable value of $\left\{\eta, 3-\eta\right\}$, to the smaller value, which is the stable solution. In this transition however, complex fields enter the stable regime at smaller field values. The additional DoF plays two roles in this transition: first by reducing the kinetic term and second by changing in the potential slope. This, however,
does not affect the extent of growth of the conventionally decaying perturbation modes, or the number of $e$-folds from the end of inflation since the time profile of Hubble parameter and the scale factor are unchanged.

We chose to work with constant $\gamma$ initially, but performed the stability analysis on $\gamma$ function space solutions, at the end. We concluded that $\gamma$=constant paths considered in the above discussions are transient for $\eta>3$. These paths are stable whenever the field is rolling down a convex potential or relaxes on an attractor. In other words, only dynamically stable trajectories in phase space are stable under $\gamma$ variations; so the potentials given in~\eqref{eqn:V(Phi)} describe the dynamically stable part of the CR complex field inflationary models, very well. We would like to stress that complex field CR models include this phenomenologically interesting subclass but are not restricted to it.

 Next, we investigated, in full generality, CR models with any number of scalar fields with non-canonical kinetic terms. For this, we used the parametrization of the field space in terms of adiabatic and the entropic directions as the ones perpendicular to it. Moreover, we explained that the procedure of finding a multifield CR model potential in separated sum form. It renders the CR model as an effective single field model with the remaining part of the field space dynamically decoupled. Including the CR constant parameter, these potentials have $N$ free parameters. The only assumption behind this general result is that the fields are minimally coupled with gravity.
\section*{Acknowledgements}

We acknowledge the financial support of the research council of the University of Tehran.



\appendix
\section{Constant Roll Potential in Canonical Coordinates}\label{sec:canonical}
For a complex field represented in terms of canonical DoFs, the Lagrangian density~\eqref{Lagrangian phi1-phi2} is very similar to that of a two-field inflationary model, in which the field equations are given by
\begin{equation}\label{KG3}
\ddot{\varphi}_1+3H\dot{\varphi_1}+\frac{\partial V}{\partial \varphi_1}=0, \qquad \ddot{\varphi}_2+3H\dot{\varphi_2}+\frac{\partial V}{\partial \varphi_2}=0.
\end{equation}
To find the CR potential in terms of canonical DoFs, we search for a relation between the DoFs such that 1) if either one of these equations in~\eqref{KG3} holds, the other equation be satisfied, as well and 2) the relation be consistent with a CR definition.

The function $Z$ introduced in~\eqref{Z definition}, written in terms of canonical DoFs, is given by $Z=\dot{\varphi}_1^2+\dot{\varphi}_2^2$. A solution in the form of $\dot{\varphi_1}=\sqrt{Z}\cos(\gamma)$ and $\dot{\varphi_2}=\sqrt{Z}\sin(\gamma)$ results in
\begin{equation}\label{linear relation}
\varphi_2=k{\varphi_1}+k_0,   \qquad  k\equiv\tan(\gamma)
\end{equation}
$k_0$ is an integration constant which can be set to zero.  
Equation~\eqref{CR constraint4} can be rewritten in terms of $\ddot{\varphi}_1$ and $\ddot{\varphi}_2$,
\begin{equation}\label{CR constraint51}
\ddot{\varphi}_1\dot{\varphi}_1+\ddot{\varphi}_2\dot{\varphi}_2=-\eta H(\dot{\varphi}_1^2+\dot{\varphi}_2^2).
\end{equation}
Note that this definition does not necessarily mean that different DoFs constant roll, independently. 
For the linear relation~\eqref{linear relation}, we have $\ddot{\varphi}_i+\eta H\dot{\varphi}_i=0$ for $i=1,2$. Furthermore, for $H=H\left(\varphi_1,\varphi_2\right)$, we will have $\dot{H}=dH/dt=\dot{\varphi_1}dH/d\varphi_1$, and the second Friedmann equation gives $-2\dot{H}=Z=(1+k^2)\dot{\varphi}_1^2$. A simple comparison relates $\dot{\varphi}_1$ to $dH/d\varphi_1$,
 \begin{equation}\label{phi dot}
\dot{\varphi}_1=-\frac{2}{1+k^2}\frac{dH}{d\varphi_1}.
\end{equation}

Plugging $\ddot{\varphi}_1=-\eta H\dot{\varphi}_1$ into the time derivative of~\eqref{phi dot} and replacing $\frac{dH}{d\varphi_1}$ in terms of $\dot{\varphi}_1$, we obtain an ordinary differential equation for $H(\varphi_1)$
\begin{equation}\label{H ODE}
\frac{d^2 H}{d\varphi_1^2}-\frac{(1+k^2)\eta}{2}H=0,
 \end{equation}
with the general solution given by
\begin{equation}\label{H solution}
H(\varphi)=C_1 e^{\sqrt{\frac{(1+k^2)\eta}{2}}\varphi_1}+C_2 e^{-\sqrt{\frac{(1+k^2)\eta}{2}}\varphi_1}.
\end{equation}
From the Friedmann equations~\eqref{Friedmann3}, we find the inflation potential for a CR complex field model as\footnote{This expression could have been found from $V=V_1(\varphi_1)+V_2(\varphi_2)$ and the relation between the fields~\eqref{linear relation}: The fields' equations of motion give us
\begin{subequations}\begin{align}\nonumber
\frac{dV_1}{d\varphi_1}&=-(3-\eta)H\dot{\varphi}_1=\frac{3-\eta}{1+k^2}\frac{d}{d\varphi_1}H^2\longrightarrow V_1=\frac{3-\eta}{1+k^2}H^2,\nonumber\\
\frac{dV_2}{d\varphi_2}&=-(3-\eta)H\dot{\varphi}_2=(3-\eta)\frac{k^2}{1+k^2}\frac{d}{d\varphi_2}H^2\longrightarrow V_2=(3-\eta)\frac{k^2}{1+k^2}H^2\nonumber.
\end{align}\end{subequations}
By adding these two terms we obtain $V=(3-\eta)\left[C_1^2 e^{\sqrt{2(1+k^2)\eta}\varphi_1}+C_2^2 e^{\sqrt{-2(1+k^2)\eta}\varphi_1}+2C_1C_2\right]$. This is equivalent (up to a constant) to~\eqref{V(phi1,phi2)-1}.  
 } 
\begin{equation}\label{V(phi1,phi2)-1}
V(\varphi_1,\varphi_2)=(3-\eta)\left[C_1^2 e^{\sqrt{2\eta(1+k^2)}\varphi_1}+C_2^2 e^{-\sqrt{2\eta(1+k^2)}\varphi_1}\right]+2C_1 C_2(3+\eta).
 \end{equation}
The one parameter CR potential that exhibits the rotational symmetry in these coordinates will, then, be given by 
\begin{equation}\label{V(phi1,phi2)}
V(\varphi_1,\varphi_2)=(3-\eta)\left[C_1^2 e^{\sqrt{2\eta(\varphi_1^2+\varphi_2^2)}}+C_2^2 e^{-\sqrt{2\eta(\varphi_1^2+\varphi_2^2)}}\right]+2C_1 C_2(3+\eta).
 \end{equation}
This is the potential solution found in \cite{Guerrero:2020lng}. There are, however subtleties in the approaches that makes the analysis in this manuscript different. The CR constraint for different fields is derived from~\eqref{CR constraint4}, whereas those are imposed by hand in \cite{Guerrero:2020lng}. Our approach is more general and can be used to any number of fields and field space metric. 
\bibliography{complex_scalar_field}{}
\bibliographystyle{JHEP}

\end{document}